\documentclass[a4paper,12pt,reqno]{article}
\usepackage{amsmath,amsfonts,amssymb,amsthm,dsfont}

\usepackage{hyperref}
\allowdisplaybreaks
\numberwithin{equation}{section}

\newcommand{\ua}{{\underline{\alpha}}}
\newcommand{\ub}{{\underline{\beta}}}
\newcommand{\ug}{{\underline{\gamma}}}
\newcommand{\ud}{{\underline{\delta}}}

\newcommand{\oa}{{\overline{\alpha}}}
\newcommand{\ob}{{\overline{\beta}}}

\newcommand{\Om}{\Omega}
\newcommand{\om}{\omega}
\newcommand{\half}{\tfrac{1}{2}}
\newcommand{\ihalf}{\tfrac{\mathrm{i}}{2}}

\newcommand{\quart}{\tfrac{1}{4}}
\newcommand{\Ii}{\mathrm{i}}

\newcommand{\4}{\tilde}

\newcommand{\6}{\partial}

\newcommand{\com}[2]{[\,#1\, ,\,#2\,]}	
\newcommand{\acom}[2]{\{#1\, ,\,#2\}}	

\newcommand{\gam}{\Gamma}	
\newcommand{\GAM}{\hat{\Gamma}}	
\newcommand{\CC}{C}	
\newcommand{\IC}{C^{-1}}	
\newcommand{\unit}{\mathds{1}}

\newcommand{\Dim}{D} 
\newcommand{\Ns}{N} 
\newcommand{\Ng}{N_\mathrm{gh}} 
\newcommand{\Omg}{\Omega_\mathrm{gh}} 
\newcommand{\brs}{s_\mathrm{susy}} 
\newcommand{\fb}{s_\mathrm{gh}} 
\newcommand{\Hg}{H_\mathrm{gh}} 
\newcommand{\rep}{F} 
\newcommand{\cdeg}{$c$-degree} 
\newcommand{\xdeg}{$\xi$-degree} 
\newcommand{\gdeg}{ghost-degree} 

\newcommand{\dds}[1]{\frac{\partial}{\partial #1}}

\newcommand{\LRA}{\Leftrightarrow}
\newcommand{\Then}{\Rightarrow}
\newcommand{\Mod}{\ \mbox{mod}\ }

\newtheorem{prop}{Proposition}[section]
\newtheorem{lemma}[prop]{Lemma}

\newcommand{\QED}{\hfill$\blacksquare$}

\begin{document}

\begin{flushright}
ITP--UH--19/09
\end{flushright}

\begin{center}
 {\large\bfseries Supersymmetry algebra cohomology I:\\[6pt] Definition and general structure}
 \\[5mm]
 Friedemann Brandt \\[2mm]
 \textit{Institut f\"ur Theoretische Physik, Leibniz Universit\"at Hannover, Appelstra\ss e 2, D-30167 Hannover, Germany}
\end{center}

\begin{abstract}
The paper concerns standard supersymmetry algebras in diverse dimensions, involving bosonic translational generators and fermionic supersymmetry generators. A cohomology related to these supersymmetry algebras, termed supersymmetry algebra cohomology, and corresponding "primitive elements" are defined by means of a BRST-type coboundary operator. A method to systematically compute this cohomology is outlined and illustrated by simple examples.
\end{abstract}

\tableofcontents

\section{Introduction}\label{sec1}

This paper is the first in a series of papers related to standard supersymmetry algebras in diverse dimensions $\Dim$. The supersymmetry algebras involve bosonic translational generators $P_a$ ($a=1,\dots,\Dim$) and fermionic supersymmetry generators $Q^i_\ua$ where $\ua$ is a spinor index and $i$ ($i=1,\dots,\Ns$) numbers sets of supersymmetry generators. The supersymmetry algebras under consideration are graded commutator algebras
 \begin{align}
  \com{P_a}{P_b}=0,\quad \com{P_a}{Q^i_\ua}=0,\quad \acom{Q^i_\ua}{Q^j_\ub}=M^{ij}(\gam^a \IC)_{\ua\ub}P_a
	\label{alg} 
 \end{align}
where $\com{A}{B}=AB-BA$ denotes the commutator of two generators $A$ and $B$ and $\acom{A}{B}=AB+BA$ denotes the anticommutator of two generators $A$ and $B$. $\gam^a$ are gamma-matrices representing a $\Dim$-dimensional Clifford algebra with signature $(t,\Dim-t)$, $\CC$ is a related charge conjugation matrix, and $M^{ij}$ are the entries of a (generally complex) $\Ns\times\Ns$ matrix (cf. sections \ref{sec2} and \ref{sec3} for details). 

The generators $P_a$, $Q^i_\ua$ are represented in some representation space $\rep$ which, in particular, may be a space of field variables of some supersymmetric field theory (cf. section \ref{1D} for examples).

The object of this work is a cohomology related to the supersymmetry algebras \eqref{alg} which we shall call supersymmetry algebra cohomology because it is an analog of Lie algebra cohomology \cite{LAC1,LAC2}. This cohomology can be suitably formulated for any supersymmetry algebra \eqref{alg} my means of a BRST-type coboundary operator $\brs$ constructed of the generators $P_a$, $Q^i_\ua$ of the supersymmetry algebra and corresponding ghost variables $c^a$ ("translation ghosts") and $\xi^\ua_i$ ("supersymmetry ghosts") according to
 \begin{align}
  \brs=c^a P_a+\xi^\ua_i \,Q_\ua^i-\half M^{ij}(\gam^a \IC)_{\ua\ub}\,\xi^\ua_i\xi^\ub_j\,\frac{\6}{\6 c^a}
	\label{brs} 
 \end{align}
where $\6/\6 c^a$ denotes an "algebraic differentiation" with respect to $c^a$ (cf. section \ref{gp}) and Einsteins summation convention is used (summation over repeated upper and lower indices; e.g., $c^a P_a$ stands for $\sum_{a=1}^\Dim c^a P_a$). The $P_a$ and $Q_\ua^i$ act on the elements of the representation space $\rep$ according to the respective representation, and trivially on the ghost variables $c^a$ and $\xi^\ua_i$ according to 
 \begin{align}
  \com{P_a}{c^b}=\acom{Q_\ua^i}{c^a}=\com{P_a}{\xi^\ua_i}=\com{Q_\ua^i}{\xi^\ub_j}=0.
	\label{PQonghosts} 
 \end{align}
Hence, denoting an element of the representation space $\rep$ by $\phi$, $\brs$ acts on this element and on the ghost variables respectively according to
 \begin{align}
  \brs\phi=c^a (P_a\phi)+\xi^\ua_i \,(Q_\ua^i\phi)\ ,\quad 
  \brs c^a=-\half M^{ij}(\gam^a \IC)_{\ua\ub}\,\xi^\ua_i\xi^\ub_j\ ,\quad 
  \brs\xi^\ua_i=0
	\label{brs1} 
 \end{align}
where $(P_a\phi)$ and $(Q_\ua^i\phi)$ are determined by the respective representation of $P_a$ and $Q_\ua^i$ in the representation space $\rep$. On functions $f(\phi,c,\xi)$ of the elements of $\rep$ and of the ghost variables, $\brs$ acts as an antiderivation (cf. section \ref{gp}).

The translation ghosts $c^a$ are fermionic (anticommuting) variables, the supersymmetry ghosts $\xi^\ua_i$ are bosonic (commuting) variables (the $c^a$ carry Grassmann parity 1, the $\xi^\ua_i$ carry Grassmann parity 0, cf. section \ref{gp}), i.e. their commutation relations read
 \begin{align}
  c^ac^b=-c^bc^a,\quad c^a\xi^\ua_i=\xi^\ua_i c^a,\quad \xi^\ua_i\xi^\ub_j=\xi^\ub_j\xi^\ua_i\ .
	\label{stat} 
 \end{align}
Equations \eqref{brs} to \eqref{stat} imply
 \begin{align*}
  \brs^2 &=\brs\circ \brs \\
  &= c^a c^bP_aP_b+c^a\xi^\ua_i (P_a Q_\ua^i-Q_\ua^i P_a)+\xi^\ua_i\xi^\ub_j Q_\ua^i Q_\ub^j
      +(\brs c^a) P_a+(\brs\xi^\ua_i) Q_\ua^i\\
  & = \half c^a c^b\com{P_a}{P_b}+c^a\xi^\ua_i \com{P_a}{Q_\ua^i}+\half \xi^\ua_i\xi^\ub_j \acom{Q_\ua^i}{Q_\ub^j}
      -\half M^{ij}(\gam^a \IC)_{\ua\ub}\,\xi^\ua_i\xi^\ub_j P_a.
 \end{align*}
The latter expression vanishes owing to the supersymmetry algebra \eqref{alg}, i.e. $\brs$ squares to zero,
 \begin{align}
  \brs^2=0.
	\label{s2} 
 \end{align}

The supersymmetry algebra cohomology is denoted by $H(\brs)$ and defined as the cohomology of $\brs$ in a space $\Om$ with members $\om(\phi,c,\xi)$ depending on the ghost variables and on the elements of $\rep$ where the dependence on the ghost variables is always polynomial and the dependence on elements of $\rep$ depends on the particular context, i.e. on the particular cohomological problem under consideration. In typical applications the dependence of members $\om\in\Om$ on elements of $\rep$ is polynomial too, but often it is further restricted by additional requirements such as an invariance of the members $\om\in\Om$ under certain transformations (e.g. an $SO(t,\Dim-t)$ invariance in the case of signature $(t,\Dim-t)$). Hence, the definition and structure of $H(\brs)$ for a supersymmetry algebra \eqref{alg} involves the representation of this algebra, i.e. the representation space $\rep$ and the way the supersymmetry algebra \eqref{alg} is represented on that space, and the particular properties of the space $\Om$ under consideration. General results concerning the dependence of $H(\brs)$ on the representations of the supersymmetry algebra \eqref{alg} seem to be unknown so far, except in the particular case $\Dim=4$, $\Ns=1$ \cite{Brandt:1991as,Brandt:1992ts,Dixon:1993yj}.

However, there is one aspect of the supersymmetry algebra cohomology which does not depend on the representation of a supersymmetry algebra \eqref{alg} at all and which we shall focus on. It is the part of the cohomology which only affects the ghost variables $c^a,\xi^\ua_i$ and involves the last part of $\brs$ in \eqref{brs}. We denote this part of $\brs$ by $\fb$,
 \begin{align}
  \fb=-\half M^{ij}(\gam^a \IC)_{\ua\ub}\,\xi^\ua_i\xi^\ub_j\,\frac{\6}{\6 c^a}\ .
	\label{fb} 
 \end{align}
Clearly, $\fb$ squares to zero by itself ($\fb^2=0$), whatever the representation of the supersymmetry algebra may be. We denote the cohomology of $\fb$ in $\Om$ by $H(\fb)$ and shall show that $H(\fb)$ can be used to systematically analyse $H(\brs)$ (cf. section \ref{general}) which is our primary motivation for studying $H(\fb)$.

Since $\fb$ only involves the ghost variables, $H(\fb)$ is obtained from the cohomology of $\fb$ in the space of polynomials in the ghost variables. We denote this space by $\Omg$ and the cohomology of $\fb$ in $\Omg$ by $\Hg(\fb)$,
 \begin{align}
  &\Omg=\Big\{\sum_{p=0}^\Dim\sum_{n=0}^r c^{a_1}\dots c^{a_p}\xi^{\ua_1}_{i_1}\dots\xi^{\ua_n}_{i_n}
  a^{i_1\dots i_n}_{\ua_1\dots\ua_n a_1\dots a_p}\,|\,
  a^{i_1\dots i_n}_{\ua_1\dots\ua_n a_1\dots a_p}\in\mathbb{C},\ r=0,1,2,\ldots\Big\},\notag\\
  &\Hg(\fb)=H(\fb,\Omg).
	\label{Omg} 
 \end{align}
We note that we do not impose any additional restriction on $\Omg$ (in particular, we shall not impose $SO(t,\Dim-t)$ invariance on the members of $\Omg$, even when the members of $\Om$ are required to be $SO(t,\Dim-t)$ invariant). Hence, in general $\Omg$ is not a subspace of $\Om$ and representatives of $\Hg(\fb)$ are not necessarily representatives of $H(\fb)$ or $H(\brs)$.
 
We term the representatives of $\Hg(\fb)$ the {\em primitive elements} of the supersymmetry algebra cohomology. According to their definition they only depend on the particular supersymmetry algebra \eqref{alg} under consideration but not on its representation.\footnote{One may interpret the primitive elements as the representatives of $H(\brs)$ for a trivial representation of the respective supersymmetry algebra \eqref{alg}, i.e. a representation with all generators $P_a, Q_\ua^i$ represented by zero and a trivial representation space, such as $\rep=\{1\}$.}

Hence, the primitive elements only depend on the spinor representation, on the number $\Ns$ of sets of supersymmetries and on the matrices $M$ and $\IC$ used in the supersymmetry algebra \eqref{alg}. The spinor representation depends on the dimension $\Dim$ and on the signature $(t,\Dim-t)$ of the Clifford algebra of the gamma-matrices.

The primary purpose of this and follow-up papers is the determination of the primitive elements of the supersymmetry algebra cohomology in various physically interesting cases (e.g. dimensions $\Dim\leq 11$, Minkowskian signature $(1,\Dim-1)$). The present paper introduces the general structures, conventions and notation for all dimensions $\Dim$ and signatures $(t,\Dim-t)$. Furthermore it explains briefly the above mentioned relation of $H(\brs)$ and $H(\fb)$ and illustrates it by simple examples.

The paper has been organized as follows. Section \ref{sec2} presents the spinor representations underlying our analysis. Section \ref{gp} introduces conventions related to the Grassmann parity. Section \ref{sec3} describes the supersymmetry algebra and the coboundary operators $\brs$ and $\fb$ in terms of Majorana type supersymmetries. In section \ref{general} general structures of $H(\fb)$ and $H(\brs)$ are discussed and related. Section \ref{1D} illustrates the structure and computation of primitive elements of a supersymmetry algebra cohomology and of $H(\brs)$ for simple examples in $\Dim=1$ dimension. Section \ref{conclusion} contains final remarks.

To the best of the authors knowledge, $\Hg(\fb)$ has been previously computed exhaustively only in the case $\Dim=4$, $\Ns=1$ for signatures $(3,1)$ and $(1,3)$ in \cite{Brandt:1991as} (see section 13.1 there) and \cite{Dixon:1993jt}. 
$H(\brs)$ has been investigated for that case for various representations of the supersymmetry algebra in spaces $\Om$ of Poincar\'e invariant functions of fields in \cite{Brandt:1991as,Brandt:1992ts,Brandt:1996au,Brandt:2002pa}, and for particular representations of the supersymmetry algebra (so-called chiral multiplets) in spaces $\Om$ of functions that need not be Poincar\'e invariant in \cite{Dixon:1993yj}.

\section{Spinor representations}\label{sec2}

In this section we present the spinor representations underlying the supersymmetry algrebras \eqref{alg} under consideration. For background concerning these spinor representations which is not reviewed here, particularly with regard to supersymmetry, we refer to \cite{VanProeyen:1999ni} and references cited there. We shall use conventions and a notation which essentially agree with those used in \cite{VanProeyen:1999ni}.\footnote{However, certain conventions and notation differ from \cite{VanProeyen:1999ni}, such as the notation and conventions for raising and lowering spinor indices, cf. equations \eqref{index3}.}

The gamma-matrices $\gam_a$ in $\Dim$ dimensions are $2^{\lfloor \Dim/2 \rfloor} \times 2^{\lfloor \Dim/2 \rfloor}$-matrices with complex entries, where $\lfloor \Dim/2 \rfloor$ denotes the largest integer less than or equal to $\Dim/2$. The gamma-matrices $\gam_a$ represent a $\Dim$-dimensional Clifford algebra with a metric $\eta_{ab}$ of signature $(-,\dots,-,+,\dots,+)$ with $t$ entries $-1$ and $\Dim-t$ entries $+1$, referred to as signature $(t,\Dim-t)$ henceforth:
 \begin{align}
  \acom{\gam_a}{\gam_b}=2\eta_{ab}\unit,\quad \eta_{ab}=
  \left\{\begin{array}{rl}-1 & \mbox{if\ $1\leq a=b\leq t$,}\\
  1& \mbox{if\ $t< a=b\leq\Dim$,}\\
  0& \mbox{if\ $a\neq b$}\end{array}\right.
	\label{cliff} 
 \end{align}
where $\unit$ denotes the $2^{\lfloor \Dim/2 \rfloor} \times 2^{\lfloor \Dim/2 \rfloor}$ unit matrix. The metric $\eta_{ab}$ and its inverse $\eta^{ab}$ are used to raise and lower the index $a$ of gamma-matrices respectively:
 \begin{align}
  \gam^a=\eta^{ab}\gam_b\ ,\quad \gam_a=\eta_{ab}\gam^b,\quad \eta^{ac}\eta_{cb}=
  \delta^a_b= \left\{\begin{array}{ll}1 & \mbox{if\ $a=b$,}\\
  0& \mbox{if\ $a\neq b$.}\end{array}\right.
	\label{raiselower} 
 \end{align}
In even dimensions $\Dim=2k$, the matrix $\GAM$ given by
 \begin{align}
  \Dim=2k:\quad 
  \GAM=(-\Ii)^{\Dim/2+t}\,\gam_1\dots\gam_\Dim
	\label{GAM} 
 \end{align}
squares to $\unit$ and anticommutes with all gamma-matrices:
 \begin{align}
  \GAM\GAM=\unit,\quad \forall a:\ \acom{\GAM}{\gam_a}=0.\label{GAMa}
 \end{align}

\subsection{Charge conjugation and complex conjugation}\label{spinors1}

The charge conjugation matrix $\CC$ relates the gamma-matrices $\gam_a$ to either the transposed gamma-matrices $\gam_a^{\top}$ or the negative transposed gamma-matrices $-\gam_a^{\top}$ according to
 \begin{align}
  \forall a:\quad\CC\,\gam_a\IC=-\eta\,\gam_a^{\top}\quad\mbox{with}\quad \eta\in\{+1,-1\},
	\label{trans} 
 \end{align}
and is either symmetric or antisymmetric,
 \begin{align}
  \CC^{\top}=-\epsilon\, \CC\quad\mbox{with}\quad \epsilon\in\{+1,-1\}.
	\label{ep} 
 \end{align}
The possible sign factors $\eta$ and $\epsilon$ depend on the dimension $\Dim$, see below. 
These sign factors determine whether the matrices $\gam^a\IC$ that occur in a supersymmetry algebra \eqref{alg} are symmetric or antisymmetric since equations \eqref{trans} and \eqref{ep} imply
 \begin{align}
  \forall a:\quad (\gam_a\IC)^{\top}=\epsilon\,\eta\, \gam_a\IC.
	\label{symm} 
 \end{align}
More generally one has for products of different gamma-matrices:
 \begin{align}
  a_i\neq a_j\ \mbox{for}\ i\neq j:\quad &(\gam_{a_1}\dots\gam_{a_k}\IC)^{\top}=
  \sigma(k)\,\gam_{a_1}\dots\gam_{a_k}\IC\notag\\
  \mbox{where}\quad
  &\sigma(k)=\left\{
  \begin{array}{rl}
  -\epsilon & \mbox{for $k\Mod 4$}\,=0,\\
  \epsilon\,\eta & \mbox{for $k\Mod 4$}\,=1,\\
  \epsilon & \mbox{for $k\Mod 4$}\,=2,\\
  -\epsilon\,\eta & \mbox{for $k\Mod 4$}\,=3.
 \end{array}
  \right.
	\label{symm2} 
 \end{align}
The gamma-matrices $\gam_a$ are related to either the conjugate-complex gamma-matrices $\gam_a^{*}$ or the negative conjugate-complex gamma-matrices $-\gam_a^{*}$ by a matrix $B$ according to
 \begin{align}
  \forall a:\quad B\,\gam_a B^{-1}=\kappa\,\gam_a^{*},\quad \kappa\in\{+1,-1\}
	\label{cc} 
 \end{align}
with a sign factor $\kappa$ depending on the respective dimension $\Dim$ and on the signature $(t,\Dim-t)$, see below. 
We choose $\CC$ and $B$ such that they are related by
 \begin{align}
   B\,C^{-1}B^\top=\kappa\,\CC^{-1*}.
   	\label{BCrel} 
 \end{align}
Owing to \eqref{trans} and \eqref{cc}, the matrix $\GAM$ defined in \eqref{GAM} fulfills
 \begin{align}
  \GAM^\top= (-1)^{\Dim/2}\CC\,\GAM \CC^{-1},\quad \GAM^*= (-1)^{\Dim/2+t}B\,\GAM B^{-1}.
	\label{GAM1} 
 \end{align}

\subsection{Standard spinor representations}\label{spinors2}

We shall now present spinor representations which we shall call standard spinor representations and which we shall use for explicit computations. These standard spinor representations are constructed in terms of the $2\times 2$ unit matrix $\sigma_0$ and the Pauli-matrices $\sigma_1,\sigma_2,\sigma_3$,
 \begin{align}
  \sigma_0=\begin{pmatrix} 1 & 0 \\ 0 & 1 \end{pmatrix} ,\quad
  \sigma_1=\begin{pmatrix} 0 & 1 \\ 1 & 0 \end{pmatrix} ,\quad
  \sigma_2=\begin{pmatrix} 0 & -\Ii \\ \Ii & 0 \end{pmatrix} ,\quad
  \sigma_3=\begin{pmatrix} 1 & 0 \\ 0 & -1 \end{pmatrix} .
	\label{sigmas} 
 \end{align}
The gamma-matrices of the standard spinor representation in $\Dim$ dimensions are defined by direct products of $\lfloor \Dim/2\rfloor$ $\sigma$-matrices according to 
 \begin{align}\begin{split}
 \Dim\in\{2k,2k+1\}:\ 
  \gam_1&=k_1\overbrace{\sigma_1\otimes \sigma_0\otimes \sigma_0\otimes\ldots\otimes\sigma_0}^{\mbox{$k$ factors}},\\
  \gam_2&=k_2\,\sigma_2\otimes \sigma_0\otimes \sigma_0\otimes\ldots\otimes\sigma_0,\\
  \gam_3&=k_3\,\sigma_3\otimes \sigma_1\otimes \sigma_0\otimes\ldots\otimes\sigma_0,\\
  \gam_4&=k_4\,\sigma_3\otimes \sigma_2\otimes \sigma_0\otimes\ldots\otimes\sigma_0,\\
  \gam_5&=k_5\,\sigma_3\otimes \sigma_3\otimes \sigma_1\otimes\ldots\otimes\sigma_0,\\
  &\ \, \vdots \\
  \gam_{2k}&=k_{2k}\,\sigma_3\otimes \sigma_3\otimes\ldots\otimes \sigma_3\otimes\sigma_2,\\
 \Dim=2k+1:\ 
  \gam_{\Dim}&=k_{\Dim}\,\sigma_3\otimes \sigma_3\otimes\ldots\otimes \sigma_3\otimes\sigma_3,
	\label{gammas} 
 \end{split}\end{align}
where $\otimes$ denotes the direct product of matrices, such as
\[
\begin{pmatrix} a_{11} & a_{12} \\ a_{21} & a_{22} \end{pmatrix}\otimes
\begin{pmatrix} b_{11} & b_{12} \\ b_{21} & b_{22} \end{pmatrix}
=
\begin{pmatrix} 
a_{11}b_{11} & a_{11}b_{12} & a_{12}b_{11} & a_{12}b_{12} \\
a_{11}b_{21} & a_{11}b_{22} & a_{12}b_{21} & a_{12}b_{22} \\
a_{21}b_{11} & a_{21}b_{12} & a_{22}b_{11} & a_{22}b_{12} \\
a_{21}b_{21} & a_{21}b_{22} & a_{22}b_{21} & a_{22}b_{22} 
\end{pmatrix},
\]
and 
\[
  k_a=\left\{\begin{array}{ll}
  \Ii & \mbox{for}\ a\leq t,\\
  1 & \mbox{for}\ a> t.
  \end{array}\right.
\]

For the corresponding matrix $\GAM$ defined in equation \eqref{GAM} one obtains, for every value of $t$:
 \begin{align}
  \Dim=2k:\quad 
  \GAM=\sigma_3\otimes\sigma_3\otimes\ldots\otimes\sigma_3\, .
	\label{GAMpart} 
 \end{align}
 

In every even dimension $\Dim=2k$ there are charge conjugation matrices $\CC$ both for $\eta=1$ and $\eta=-1$ (this reflects that in even dimensions a set of gamma-matrices $\{\gam_1,\ldots,\gam_\Dim\}$ is equivalent to the set $\{-\gam_1,\ldots,-\gam_\Dim\}$ owing to $\gam_a=-\GAM\gam_a\GAM$). In the standard spinor representations these charge conjugation matrices $\CC$ are chosen according to
 \begin{align}
  \Dim=2k:\quad\CC=\left\{\begin{array}{ll}
  \sigma_2\otimes \sigma_1\otimes \sigma_2\otimes\sigma_1\otimes\ldots & \mbox{for}\ 
  \eta=+1,\\
  \sigma_1\otimes \sigma_2\otimes \sigma_1\otimes\sigma_2\otimes\ldots& \mbox{for}\ 
  \eta=-1.
  \end{array}\right.
	\label{C1} 
 \end{align}
In an odd dimension $\Dim=2k+1$ only one of the matrices in equation \eqref{C1} fulfills equation \eqref{trans} (this reflects that in odd dimensions a set of gamma-matrices $\{\gam_1,\ldots,\gam_\Dim\}$ is not equivalent to the set $\{-\gam_1,\ldots,-\gam_\Dim\}$). Indeed, $\gam_{\Dim}$ in $\Dim=2k+1$ dimensions is proportional to $\GAM$ in $\Dim=2k$ dimensions and thus fulfills $\gam_\Dim^\top= (-1)^{k}\CC\gam_\Dim \IC$ according to the first equation \eqref{GAM1}. Hence, in order to fulfill equation \eqref{trans} in $\Dim=2k+1$ dimensions with a matrix $\CC$ as in equation \eqref{C1}, the sign factor $\eta$ must fulfill $-\eta= (-1)^{k}$. This yields:
\begin{align}
  \Dim=2k+1:\quad\CC=\left\{\begin{array}{lll}
  \sigma_2\otimes \sigma_1\otimes \sigma_2\otimes\sigma_1\otimes\ldots & \mbox{for}\ \Dim\ \mbox{mod}\ 4=3 &
  (\eta=+1),\\
  \sigma_1\otimes \sigma_2\otimes \sigma_1\otimes\sigma_2\otimes\ldots& \mbox{for}\ \Dim\ \mbox{mod}\ 4=1 &
  (\eta=-1).
  \end{array}\right.
	\label{C2} 
\end{align}
The respective values of $\epsilon$ are easily obtained from equations \eqref{C1} and \eqref{C2} for the various cases.

For the matrix $B$ there are two options in any standard spinor representation. The first option is $B=B_{(1)}$ with
 \begin{align}
  B_{(1)}=b_{(1)}\left\{\begin{array}{ll}
  \CC\, \gam_1\dots\gam_t & 
  \mbox{for}\ t>0,\\
  \CC&
  \mbox{for}\ t=0,
  \end{array}\right.
	\label{B1} 
 \end{align}
where, in order to fulfill equation \eqref{BCrel}, $b_{(1)}$ is chosen such that
 \begin{align}
  (b_{(1)})^2=(-\eta)^{t+1}.
	\label{b1} 
 \end{align}
$B_{(1)}$ fulfills equation \eqref{cc} with $\kappa=- (-1)^t\eta$:
 \begin{align}
  \forall a:\quad B_{(1)}\,\gam_a B_{(1)}^{-1}=- (-1)^t\eta\,\gam_a^{*}.
	\label{cc1} 
 \end{align}

The second option is $B=B_{(2)}$ with 
 \begin{align}
  B_{(2)}=b_{(2)}\left\{\begin{array}{ll}
  \CC\, \gam_{t+1}\dots\gam_\Dim & 
  \mbox{for}\ t<\Dim,\\
  \CC&
  \mbox{for}\ t=\Dim,
  \end{array}\right.
	\label{B2} 
 \end{align}
where, in order to fulfill equation \eqref{BCrel}, $b_{(2)}$ is chosen such that
 \begin{align}
  (b_{(2)})^2=\eta^{\Dim-t+1}.
	\label{b2} 
 \end{align}
$B_{(2)}$ fulfills equation \eqref{cc} with $\kappa= (-1)^{\Dim-t}\eta$:
 \begin{align}
  \forall a:\quad B_{(2)}\,\gam_a B_{(2)}^{-1}= (-1)^{\Dim-t}\eta\,\gam_a^{*}.
	\label{cc2} 
 \end{align}
Notice that the sign factors in equations \eqref{cc1} and \eqref{cc2} differ in even dimensions and agree in odd dimensions (again, this reflects that the sets $\{\gam_1,\ldots,\gam_\Dim\}$ and $\{-\gam_1,\ldots,-\gam_\Dim\}$ are equivalent in even dimensions but inequivalent in odd dimensions).

In $\Dim=1$ dimension we use
 \begin{align}
  \Dim=t=1:\quad\gam_1=\Ii,\quad \CC=1,\quad B=\Ii.
	\label{1D0} 
 \end{align}
 
\subsection{Spinor indices}\label{spinors3}

The entries of the gamma-matrices $\gam_a$, transposed gamma-matrices $\gam_a^{\top}$, conjugate-complex gamma-matrices $\gam_a^{*}$ and adjoint gamma-matrices $\gam_a^{\dagger}=\gam_a^{\top *}$ are denoted by, respectively,
 \begin{align}
  \gam_{a\,\ua}{}^\ub\, ,\quad \gam_a^{\top\ub}{}_\ua\, ,\quad \gam_a^{*}{}_\oa{}^\ob\, ,\quad \gam_a^{\dagger\ob}{}_\oa
	\label{index1} 
 \end{align}
where in each case the left spinor index (whether up or down) numbers the rows and the right spinor index numbers the columns of the respective matrix. As in \eqref{index1}, complex conjugation of an object is indicated by a star $*$ and by interchanging underlining and overlining of spinor indices.\footnote{The position (up or down) and underlining or overlining of spinor indices indicate the transformation properties under $\mathfrak{so}(t,\Dim-t)$ transformations, cf. section \ref{spinors6a}.} The corresponding index structure of the matrices $\CC$, $B$ and the respective inverted, conjugate-complex and conjugate-complex inverted matrices is
 \begin{align}
  \CC^{\ua\ub},\quad \CC^{-1}{}_{\ua\ub}\, ,\quad \CC^{*\oa\ob},\quad \CC^{-1*}{}_{\oa\ob}\, ,\quad
  B_\oa{}^\ub,\quad B^{-1}{}_\ua{}^\ob,\quad B^*{}_\ua{}^\ob,\quad B^{-1*}{}_\oa{}^\ub
	\label{index2} 
 \end{align}
where, again, the left spinor index is a row index and the right spinor index is a column index, respectively.
The charge conjugation matrix $\CC$, its inverse $\IC$ and the corresponding conjugate-complex matrices $\CC^*$ and $\CC^{-1*}$ are used to raise and lower indices of spinors $\psi_\ua$, $\psi^\ua$, $\psi_\oa$ and $\psi^\oa$, respectively, 
according to
 \begin{align}
  \psi^\ua=\CC^{\ua\ub}\psi_\ub\, ,\quad \psi_\ua=\IC{}_{\ua\ub}\psi^\ub,\quad
  \psi^\oa=\CC^{*\oa\ob}\psi_\ob\, ,\quad \psi_\oa=\CC^{-1*}{}_{\oa\ob}\psi^\ob.
	\label{index3} 
 \end{align}
We remark that raising and lowering of the spinor indices of the gamma-matrices and of the conjugate-complex gamma-matrices (and, analogously, of the matrices $\CC$, $B$ etc.) must not be confused with the transposition of these matrices. E.g., 
\[
\gam_a{}^{\ub}{}_\ua=\CC^{\ub\ug}\IC{}_{\ua\ud}\gam_{a\,\ug}{}^\ud=-\epsilon\,(\CC\gam_a\IC)^\ub{}_\ua
\] 
is in general different from 
\[
\gam_a^{\top\ub}{}_\ua=-\eta\, (\CC\gam_a\IC)^\ub{}_\ua\, . 
\]

\subsection{Majorana and symplectic Majorana spinors}\label{spinors4} 

Majorana spinors $\psi_\ua$ with lower spinor indices $\ua$ are defined by the requirement that they are related to the conjugate-complex spinors through the matrix $B$ according to $\psi^*_\oa=bB_\oa{}^\ub\psi_\ub$ with some phase factor $b$, $|b|=1$, that can be chosen appropriately (and depending on the respective spinor $\psi$). This requires $B^*B=\unit$ owing to
 \begin{align*} 
  \psi^*_\oa=bB_\oa{}^\ub\psi_\ub\,,\ |b|=1\ \Rightarrow\ \psi_\ua&=(\psi^{*}_\oa)^*=(bB_\oa{}^\ub\psi_\ub)^*= 
  b^*B^*{}_\ua{}^\ob\psi^*_\ob\\
  &=b^*B^*{}_\ua{}^\ob bB_\ob{}^\ug\psi_\ug=(B^*B)_\ua{}^\ub\psi_\ub\, .
 \end{align*}
The two options \eqref{B1} and \eqref{B2} for $B$ yield, respectively:
 \begin{align} 
  B_{(1)}^*B_{(1)}&=-\epsilon\,\eta^t\, (-1)^{t(t+1)/2}\,\unit,
	\label{B*B1} \\
	B_{(2)}^*B_{(2)}&=-\epsilon\,\eta^{\Dim-t}\, (-1)^{(\Dim-t)(\Dim-t+1)/2}\,\unit.
	\label{B*B2}
 \end{align}
Hence, in general it depends on the dimension $\Dim$ and on the signature $(t,\Dim-t)$ whether and how Majorana spinors can be defined. For upper indices of Majorana spinors one obtains:
 \begin{align*}
  \psi^*_\oa=bB_\oa{}^\ub\psi_\ub\ \Rightarrow\ 
  \psi^{*\oa}
  &=\CC^{*\oa\ob}\psi^*_\ob=b(\CC^*B)^{\oa\ub}\psi_\ub=b(\CC^*B\IC)^\oa{}_\ub\psi^\ub\\
  &=b\kappa B^{-1\top\,\oa}{}_\ub\psi^\ub=b\kappa\,\psi^\ub B^{-1}{}_\ub{}^\oa
 \end{align*}
where we have used \eqref{BCrel}.
Hence we define Majorana spinors with lower and upper spinor indices according to:
 \begin{align}
  \mbox{Majorana spinors ($B^*B=\unit$):}\quad
  \psi^*_\oa=bB_\oa{}^\ub\psi_\ub\, ,\quad \psi^{*\oa}=b\kappa\,\psi^\ub B^{-1}{}_\ub{}^\oa,\quad |b|=1.
	\label{MS} 
 \end{align}
 
If $B^*B=-\unit$ Majorana spinors do not exist. Nevertheless, one can still impose a reality condition on spinors $\psi^i$, a so-called "symplectic Majorana condition", when there are at least two sets of spinors ($\Ns\geq 2$):
 \begin{align}
  &\mbox{symplectic Majorana spinors ($B^*B=-\unit$):}\notag\\
  &\psi^{*}_{i\,\oa}=\Omega_{ij}B_\oa{}^\ub\psi^j_\ub\, ,\quad
   \psi^{*\,\oa}_i=\kappa\,\Omega_{ij}\psi^{j\,\ub}B^{-1}{}_\ub{}^\oa,
	\label{SMS1} \\
  &\psi^{*i\,\oa}=-\psi^{\ub}_j B^{-1}{}_\ub{}^\oa\Omega^{*ji},\quad
  \psi^{*i}_{\oa}=-\kappa\,B_\oa{}^\ub\psi_{j\,\ub}\Omega^{*ji} 
	\label{SMS1a} 
 \end{align}
where we have used the convention that complex conjugation changes the position (up or down) of an $i$-index,
 \begin{align} 
  \psi^{*}_{i\,\oa}:=(\psi^i_\ua)^*,\quad
  \psi^{*i}_{\oa}:=(\psi_{i\,\ua})^*\, ,
	\label{complexi} 
 \end{align}
and $\Omega_{ij}$ are (in general complex) entries of an invertible matrix $\Omega$ fulfilling $\Omega^{-1}=-\Omega^*$:
 \begin{align}
   \Omega^{*ik}\Omega_{kj}=-\delta^i_j\quad\mbox{with}\quad  \Omega^{*ij}:=(\Omega_{ij})^*.
	\label{SMS2} 
 \end{align}
\eqref{SMS2} is required by consistency: e.g., for $\psi^i_\ua$ one obtains $\psi^i=(\psi_i^*)^*=(\Omega_{ij}B\psi^j)^*=\Omega^{*ij}B^*\psi^*_j=\Omega^{*ij}\Omega_{jk}B^*B\psi^k=-\Omega^{*ij}\Omega_{jk}\psi^k$ where in the last step we used $B^*B=-\unit$.

\subsection{Majorana-Weyl and symplectic Majorana-Weyl spinors}\label{spinors5}

In even dimensions, Weyl spinors (chiral spinors) are eigenspinors of $\GAM$. Owing to $\GAM^2=\unit$, $\GAM$ only has eigenvalues $+1$ and $-1$:
 \begin{align}
  \mbox{Weyl spinors ($\Dim=2k$):}\quad 
  \psi_\ua=c\,\GAM_\ua{}^\ub\psi_\ub\,,\quad c\in\{+1,-1\}.
	\label{Weyl} 
 \end{align}
The eigenvalue $c$ will be called "chirality" of a Weyl spinor. 

Corresponding Weyl spinors with upper indices $\ua$ fulfill accordingly
\[
\psi^\ua=\CC^{\ua\ub}\psi_{\ub}=c\,(\CC\,\GAM)^{\ua\ub}\psi_{\ub}
=c\,(\CC\,\GAM\IC)^\ua{}_\ub\psi^\ub=c\, (-1)^{D/2}\psi^\ub\GAM_\ub{}^\ua
\]
where we have used the first equation \eqref{GAM1}. Weyl spinors with overlined spinor indices are accordingly eigenspinors of $\GAM^*$.

Majorana-Weyl spinors are Weyl spinors fulfilling the Majorana condition \eqref{MS}. Symplectic Majorana-Weyl spinors are Weyl spinors fulfilling the symplectic Majorana condition \eqref{SMS1} (with $\Omega$ fulfilling \eqref{SMS2}). Such spinors only exist if $B\,\GAM=\GAM^*B$. The latter condition arises because, e.g., consistency requires for a Majorana-Weyl spinor $\psi_{\ua}$ that $\psi^*=bB\psi=bc\, B\,\GAM\psi$ and $\psi^*=(c\,\GAM\psi)^*=c\,\GAM^*\psi^*=
bc\,\GAM^*B\psi$ are equal. The second equation \eqref{GAM1} yields
 \begin{align}
 \GAM^*B= (-1)^{\Dim/2+t}B\GAM B^{-1}B= (-1)^{\Dim/2+t}B\GAM.
 \label{MW1}
 \end{align}
Hence, Majorana-Weyl spinors only exist in even dimensions when both $ (-1)^{\Dim/2+t}=1$ and $B^*B=\unit$ hold.
Symplectic Majorana-Weyl spinors only exist in even dimensions when both $ (-1)^{\Dim/2+t}=1$ and $B^*B=-\unit$ hold.

\subsection{\texorpdfstring{$\mathfrak{so}(t,\Dim-t)$ transformations}{so(t,D-t) transformations}}\label{spinors6a}

The supersymmetry algebra \eqref{alg} is for signature $(t,\Dim-t)$ form-invariant under transformations which form the Lie algebra $\mathfrak{so}(t,\Dim-t)$. This Lie algebra will be helpful to describe the supersymmetry algebra cohomology in $\Dim$ dimensions for the signature $(t,\Dim-t)$ and therefore will be briefly introduced in the following.
We denote the real generators of the Lie algebra $\mathfrak{so}(t,\Dim-t)$ by $\ell_{ab}$ (with $\ell_{ab}=-\ell_{ba}$) and choose a basis of these generators such that their commutator algebra reads
 \begin{align}
  \com{\ell_{ab}}{\ell_{cd}}=\eta_{ad}\ell_{bc}-\eta_{ac}\ell_{bd}-\eta_{bd}\ell_{ac}+\eta_{bc}\ell_{ad}\, .
	\label{Lor1} 
 \end{align}
The generators $\ell_{ab}$ are represented on $\mathfrak{so}(t,\Dim-t)$-covariant vectors with components $v_a$, $\mathfrak{so}(t,\Dim-t)$-contravariant vectors with components $v^a$ and spinors with components $\psi_\ua$, $\psi^\ua$, $\psi_\oa$ and $\psi^\oa$ respectively according to
 \begin{align}
  &\ell_{ab}v_c=(\eta_{bc}\delta_a^d-\eta_{ac}\delta_b^d)v_d\, ,\quad
   \ell_{ab}v^c=(\delta_b^c\eta_{ad}-\delta_a^c\eta_{bd})v^d,\notag\\
  &\ell_{ab}\psi_\ua=-\Sigma_{ab\,\ua}{}^\ub\psi_\ub\, ,\quad
   \ell_{ab}\psi^\ua=\psi^\ub\Sigma_{ab\,\ub}{}^\ua ,\notag\\
  &\ell_{ab}\psi_\oa=-\Sigma^*_{ab\,\oa}{}^\ob\psi_\ob\, ,\quad 
   \ell_{ab}\psi^\oa=\psi^\ob\Sigma^*_{ab\,\ob}{}^\oa
	\label{Lor2} 
 \end{align}
where
 \begin{align}
  \Sigma_{ab\,\ua}{}^\ub=\quart\com{\gam_a}{\gam_b}_\ua{}^\ub\,,\quad 
  \Sigma^*_{ab\,\oa}{}^\ob=(\Sigma_{ab\,\ua}{}^\ub)^*=(B\Sigma_{ab} B^{-1})_\oa{}^\ob\, .
	\label{Lor3} 
 \end{align}
We denote $\mathfrak{so}(t,\Dim-t)$-invariant products of spinors by a dot-symbol and define them for spinors with indices $\ua,\,\ub$ according to
 \begin{align}
  \psi\cdot\chi:=\psi^\ua\chi^\ub\,\IC_{\ua\ub}=-\epsilon\,\psi_\ua\chi_\ub \,\CC^{\ua\ub}\,.
	\label{Lor4} 
 \end{align}
The corresponding $\mathfrak{so}(t,\Dim-t)$-invariant products of spinors with spinor indices $\oa,\,\ob$ follow from \eqref{Lor4} by complex conjugation using the rules given in section \ref{gp}.

\subsection{Equivalent spinor representations}\label{spinors6}

Two sets $\{\gam^a\}$, $\{\gam^{\prime\,a}\}$ of gamma-matrices are called equivalent if they are related by an invertible (not necessarily unitary) complex matrix $R$ according to
 \begin{align}
  \forall a:\quad \gam^{\prime\,a}=R\,\gam^a R^{-1}.
	\label{eq1} 
 \end{align}
When passing in this way from one set of gamma-matrices to an equivalent set, we simultaneously pass from the matrices $\CC$ and $B$ to matrices $\CC^{\,\prime}$ and $B^{\,\prime}$ given by
 \begin{align}
  \CC^{\,\prime}=R^{-1\top}\CC R^{-1},\quad B^{\,\prime}=R^*BR^{-1}.
	\label{eq2} 
 \end{align}
One readily checks that \eqref{eq1} and \eqref{eq2} preserve the above equations \eqref{trans} to \eqref{GAM1} as well as \eqref{B*B1} and \eqref{B*B2} (with unchanged values of $\eta$, $\epsilon$ and $\kappa$) in the sense that the latter equations hold for the primed matrices whenever they hold for unprimed matrices.
Furthermore, the Majorana condition \eqref{MS}, the symplectic Majorana condition \eqref{SMS1} and the Weyl condition \eqref{Weyl} are preserved (without changing the matrix $\Omega$ in equation \eqref{SMS1}, and with $\GAM^{\,\prime}=R\,\GAM R^{-1}$ in even dimensions), when we relate spinors accordingly by
 \begin{align}
  \psi^{\prime}_\ua=R_\ua{}^\ub \psi_\ub\, ,\quad \psi^{\prime\,\ua}=\psi^{\ub}R^{-1}{}_\ub{}^\ua
	\label{eq3} 
 \end{align}
and the respective conjugate-complex relations.
In particular, using equations \eqref{eq1} to \eqref{eq3}, one can pass from a standard spinor representation of the $\gam^a$, $\CC$ and $B$ given section \ref{spinors2} to any equivalent spinor representation, keeping all the features \eqref{trans} to \eqref{GAM1} as well as the respective (possibly symplectic) Majorana or Majorana-Weyl condition.

A consequence of the above relations between equivalent spinor representations in even dimensions $\Dim$ is that they do not mix chiralities in the following sense: when passing from a first spinor representation to an equivalent second spinor representation, the components of a spinor with positive chirality in the second spinor representation are always linear combinations of the components of the corresponding spinor with the same chirality in the first spinor representation, and analogously for spinors with negative chirality,
 \begin{align}
 &\psi_\ua=c\,\GAM_\ua{}^\ub\psi_\ub\quad \Then\quad \psi^\prime_\ua=c\,\GAM^{\,\prime}_\ua{}^\ub\psi^\prime_\ub\,;\notag\\
 &\psi^\ua=c\,\psi^\ub\GAM_\ub{}^\ua\quad \Then\quad \psi^{\prime\,\ua}=c\,\psi^{\prime\,\ub}\GAM^{\,\prime}_\ub{}^\ua .
	\label{eq4} 
 \end{align}
This applies, in particular, to spinors $\psi$ composed polynomially in an $\mathfrak{so}(t,\Dim-t)$-covariant manner of the ghost variables $c^a$ and $\xi^\ua_i$, such as $c^a\xi^\ua_i\gam_{a\,\ua}{}^\ub$. 

\subsection{Summary of features of spinor representations}\label{spinors7}

We shall now summarize features of spinor representations presented above which depend on the dimension $\Dim$ and on the signature $(t,\Dim-t)$. Table \eqref{table2} collects the possible values of $\eta$ and $\epsilon$ in the various dimensions $\Dim$ and indicates, depending on the signature $(t,\Dim-t)$, whether there are Majorana-Weyl spinors ($MW$), just Majorana spinors ($M$), symplectic Majorana-Weyl spinors ($SMW$) or just symplectic Majorana spinors ($SM$). The table also indicates whether a matrix $B_{(1)}$ or $B_{(2)}$ (given in equations \eqref{B1}, \eqref{B2} for the standard spinor representations) may be used to define Majorana, Majorana-Weyl or symplectic Majorana-Weyl spinors in the various cases which allow such spinors. To that end, the respective matrix ($B_{(1)}$ or $B_{(2)}$) is given in parantheses if only that matrix can be used in a particular case; if no matrix is given in parantheses, one may use either $B_{(1)}$ or $B_{(2)}$. In fact, the cases in which the choice of $B_{(1)}$ or $B_{(2)}$ matters are those for which $\Dim$ is even and $\Dim/2+t$ is odd (as these are the only cases for which the sign factors on the right hand sides of \eqref{B*B1} and \eqref{B*B2} differ). These are precisely those cases in even dimensions for which Majorana spinors but no Majorana-Weyl spinors exist. All the features depend modulo 8 on the dimension $\Dim$ and modulo 4 on the value of $t$.
 \begin{align}
 \begin{array}{|c|c|c|c|c|c|c|}
 \hline
 \Dim\ \mbox{mod}\ 8 &\eta & \epsilon &t\ \mbox{mod}\ 4=0 &t\ \mbox{mod}\ 4=1 &t\ \mbox{mod}\ 4=2 &t\ \mbox{mod}\ 4=3\\
 \hline
 0 & +1 & -1 & MW & M\ (B_{(2)}) & SMW & M\ (B_{(1)})\\
 \hline
 0 & -1 & -1 & MW & M\ (B_{(1)}) & SMW & M\ (B_{(2)})\\
 \hline
 1 & -1 & -1 & M & M & SM & SM\\
 \hline
 2 & +1 & +1 & M\ (B_{(2)}) & MW & M\ (B_{(1)}) & SMW\\
 \hline
 2 & -1 & -1 & M\ (B_{(1)}) & MW & M\ (B_{(2)}) & SMW\\
 \hline
 3 & +1 & +1 & SM & M & M & SM\\
 \hline
 4 & +1 & +1 & SMW & M\ (B_{(1)}) & MW & M\ (B_{(2)}) \\
 \hline
 4 & -1 & +1 & SMW & M\ (B_{(2)}) & MW & M\ (B_{(1)}) \\
 \hline
 5 & -1 & +1 & SM & SM & M & M \\
 \hline
 6 & +1 & -1 & M\ (B_{(1)}) & SMW & M\ (B_{(2)}) & MW\\
 \hline
 6 & -1 & +1 & M\ (B_{(2)}) & SMW & M\ (B_{(1)}) & MW\\
 \hline
 7 & +1 & -1 & M & SM & SM & M\\
 \hline
 \end{array}
	\label{table2} 
 \end{align}

Table \eqref{table2} shows in particular that in all dimensions $\Dim$ the signatures $(t,\Dim-t)$ and $(\Dim-t,t)$ have corresponding properties concerning reality properties of spinors in the sense that corresponding types of spinors ($MW$, $M$, $SMW$ or $SM$) exist (however, as remarked above, the matrix $B$ needed to define such spinors may differ for the signatures $(t,\Dim-t)$ and $(\Dim-t,t)$). This reflects that these signatures are related just by changing the overall sign of the metric $\eta_{ab}$. 
 
\section{Grassmann parity and related conventions}\label{gp}

The Grassmann parity is an attribute to describe algebraic features of the objects relevant to the supersymmetry algebra \eqref{alg} and the related cohomology $H(\brs)$. These objects are "variables", and "operators" acting on (functions of) the variables. 

The "variables" are the elements of the representation space $\rep$, the translation ghosts $c^a$ and the supersymmetry ghosts $\xi^\ua_i$. The variables are treated as algebraically independent objects, possibly modulo algebraic relations such as Majorana oder symplectic Majorana conditions. However, when such algebraic conditions between variables are present one alternatively may work with a smaller number of variables that are algebraically independent. E.g., the components of a Majorana spinor $\psi$ (which is not subject to any further condition apart from the Majorana condition) and the conjugate-complex spinor $\psi^*$ are not algebraically independent as they are related by the Majorana condition \eqref{MS}. Hence, one may take the components of $\psi$ or $\psi^*$ as algebraically independent variables, but not both of them simultaneously.

"Operators" are, in particular, the generators $P_a$, $Q^i_\ua$ occurring in supersymmetry algebra \eqref{alg} and the coboundary operators $\brs$ and $\fb$ defined in equations \eqref{brs} and \eqref{fb}.

The Grassmann parity of an object $X$ is denoted by $|X|\in\{0,1\}$. The Grassmann parities of the variables determine their commutation relations according to
 \begin{align}
  \varphi^1\varphi^2= (-1)^{|\varphi^1|\,|\varphi^2|}\varphi^2\varphi^1,\quad \varphi^1,\varphi^2\in\{c^a,\xi^\ua_i\}\cup\rep.
	\label{gp1} 
 \end{align}
The translation ghosts $c^a$ are Grassmann odd, the supersymmetry ghosts $\xi^\ua_i$ are Grassmann even variables,
 \begin{align}
  |c^a|=1,\quad |\xi^\ua_i|=0.
	\label{gp2} 
 \end{align}
The Grassmann parity is additive modulo 2 in the sense that
 \begin{align}
  |\varphi^1\varphi^2\dots\varphi^m|=(|\varphi^1|+|\varphi^2|+\dots+|\varphi^m|)\ \mbox{mod 2}.
	\label{gp3} 
 \end{align}
A first order differential operator $\gamma$ with Grassmann parity $|\gamma|\in\{0,1\}$ satisfies a graded Leibniz rule on products of variables: a derivation is Grassmann even and an antiderivation is Grassmann odd,
  \begin{align}     \gamma(\varphi^1\varphi^2)=(\gamma\varphi^1)\varphi^2+ (-1)^{|\gamma|\,|\varphi^1|}\varphi^1(\gamma\varphi^2).
	\label{gp4} 
 \end{align}
The translational generators are derivations, the supersymmetry generators $Q_\ua^i$ and the coboundary operators $\brs$ and $\fb$ are antiderivations,
  \begin{align}
  |P_a|=0,\quad |Q_\ua^i|=|\brs|=|\fb|=1.
	\label{gp5} 
 \end{align}
A particular first order differential operator is the "algebraic differentiation" $\6/\6\varphi$ with respect to a variable $\varphi$. It is defined according to 
\begin{align}
\frac{\6}{\6\varphi^1}\,\varphi^2=\frac{\6\varphi^2}{\6\varphi^1}+ (-1)^{|\varphi^1|\,|\varphi^2|}\varphi^2\,\frac{\6}{\6\varphi^1}\ ,\quad
\frac{\6\varphi^2}{\6\varphi^1}=\left\{\begin{array}{rl}
  1 & \mbox{if}\ \varphi^1=\varphi^2,\\
  0 & \mbox{if}\ \varphi^1\neq\varphi^2.
  \end{array}\right.
\label{gp6} 
\end{align}
In particular, $\6/\6\varphi$ thus has the same Grassmann parity as $\varphi$,
  \begin{align}
  \left|\frac{\6}{\6\varphi}\right|=|\varphi|.
	\label{gp7} 
 \end{align}

Complex conjugation is defined with a sign factor depending on the Grassmann parity: the conjugate-complex of the product of two objects (variables and/or operators) is defined as the product of the conjugate-complex objects times a sign factor which is negative when both objects are Grassmann odd and positive otherwise,
  \begin{align}
  (XY)^*= (-1)^{|X|\,|Y|}X^*Y^*.
	\label{gp8} 
 \end{align}
This implies, for instance, that
the conjugate-complex $(\gamma\varphi)^*$ of a real operator $\gamma$ acting on a real variable $\varphi$ equals $-\gamma\varphi$ when both $\gamma$ and $\varphi$ are Grassmann odd.

One infers from equations \eqref{gp6}, \eqref{gp7} and \eqref{gp8} that the conjugate-complex of the algebraic differentiation $\6/\6\varphi$ with respect to a variable $\varphi$ is the algebraic differentiation $\6/\6\varphi^*$ with respect to the conjugate-complex variable $\varphi^*$ times the sign factor $ (-1)^{|\varphi|}$, 
 \begin{align}
  \left(\frac{\6}{\6\varphi}\right)^*= (-1)^{|\varphi|}\,\frac{\6}{\6\varphi^*}\ .
	\label{gp9} 
 \end{align}
In particular, the algebraic differentiation $\6/\6\varphi$ with respect to a real variable $\varphi$ is thus a purely imaginary differential operator when $\varphi$ is Grassmann odd,
 \begin{align}
  \varphi=\varphi^* \Rightarrow\ \left(\frac{\6}{\6\varphi}\right)^*= (-1)^{|\varphi|}\,\frac{\6}{\6\varphi}\ .
	\label{gp10} 
 \end{align}
The translation ghosts $c^a$ and the translational generators are real,
 \begin{align}
  (c^a)^*=c^a,\quad (P_a)^*=P_a.
	\label{gp11} 
 \end{align}
The reality properties of the supersymmetry generators $Q_\ua^i$, of the supersymmetry ghosts $\xi^\ua_i$ and of the coboundary operators $\brs$ and $\fb$ are discussed in section \ref{sec3}.

Notice that equations \eqref{gp2}, \eqref{gp10} and \eqref{gp11} imply that the algebraic differentiations $\6/\6c^a$ that occur in the definitions \eqref{brs}, \eqref{fb} of $\brs$ and $\fb$ are purely imaginary operations,
 \begin{align}
  \left(\frac{\6}{\6c^a}\right)^*=-\,\frac{\6}{\6c^a}\ .
	\label{gp12} 
 \end{align}

\section{Majorana type supersymmetries}\label{sec3}

We shall formulate and investigate the supersymmetry algebra cohomology in terms of Majorana type supersymmetries. In this section we introduce the corresponding structures and notation.

\subsection{Supersymmetry algebra}\label{MSusy}

Supersymmetry generators $Q_\ua^i$ of 
Majorana supersymmetries fulfill a Majorana condition \eqref{MS}. With no loss of generality we define
 \begin{align}
  \mbox{Majorana supersymmetries ($B^*B=\unit$):}\quad Q^*_{i\,\oa}=B_\oa{}^\ub Q^i_\ub\quad
  \mbox{with}\quad Q^*_{i\,\oa}:=(Q_\ua^i)^*.
	\label{MSusy1} 
 \end{align}
This implies a reality condition on the matrix $M$ that occurs in the anticommutators $\acom{Q^i_\ua}{Q^j_\ub}$ of the supersymmetry algebra \eqref{alg}. Indeed, \eqref{MSusy1} implies
 \begin{align}\begin{split}
  (\acom{Q^i_\ua}{Q^j_\ub})^*&=-\acom{Q^*_{i\,\oa}}{Q^*_{j\,\ob}}=
  -B_\oa{}^\ug B_\ob{}^\ud\acom{Q^i_\ug}{Q^j_\ud}\\
  &=
  -B_\oa{}^\ug B_\ob{}^\ud M^{ij}(\gam^a\IC)_{\ug\ud} P_a
	\label{MSusy2} 
 \end{split}\end{align}
where the minus sign originates from equations \eqref{gp5} and \eqref{gp8}.
According to \eqref{alg}, \eqref{MSusy2} must be equal to
 \begin{align}\begin{split}
  (M^{ij}(\gam^a\IC)_{\ua\ub}P_a)^*&=M^*_{ij}(\gam^{a*}\CC^{-1*})_{\oa\ob}P_a
  =M^*_{ij}(B\,\gam^a B^{-1}B\IC B^\top)_{\oa\ob}P_a\\
  &=
  B_\oa{}^\ug B_\ob{}^\ud M^*_{ij}(\gam^a\IC)_{\ug\ud} P_a
	\label{MSusy3} 
 \end{split}\end{align}
where we used equations \eqref{cc}, \eqref{gp11}, \eqref{BCrel} and the notation
 \begin{align}
  M^*_{ij}:=(M^{ij})^*.
	\label{MSusy4} 
 \end{align}
We read off from equations \eqref{MSusy2} and \eqref{MSusy3} that for Majorana supersymmetries \eqref{MSusy1} the entries of $M$ are purely imaginary:
 \begin{align}
  \mbox{for Majorana supersymmetries:}\quad M^*_{ij}=-M^{ij}.
	\label{MSusy5} 
 \end{align}

The supersymmetry generators $Q_\ua^i$ of symplectic
Majorana supersymmetries fulfill a symplectic Majorana condition \eqref{SMS1},
 \begin{align}
  \mbox{symplectic Majorana supersymmetries ($B^*B=-\unit$):}\quad Q^*_{i\,\oa}=\Omega_{ij} B_\oa{}^\ub Q^j_\ub
	\label{MSusy9} 
 \end{align}
where $\Omega$ is subject to \eqref{SMS2}.
Analogously to \eqref{MSusy2} one derives from \eqref{MSusy9} that
 \begin{align}
  (\acom{Q^i_\ua}{Q^j_\ub})^*=
  -\Omega_{ik}\Omega_{jl}B_\oa{}^\ug B_\ob{}^\ud M^{kl}(\gam^a\IC)_{\ug\ud} P_a\ .
	\label{MSusy10} 
 \end{align}
The latter expression must be equal to the expression \eqref{MSusy3}. This yields
 \begin{align}
  \mbox{for symplectic Majorana supersymmetries:}\quad M^*_{ij}=-\Omega_{ik}\Omega_{jl}M^{kl}.
	\label{MSusy11} 
 \end{align}

We recall that the product $\epsilon\,\eta$ determines the symmetry of the matrices $\gam^a\IC$ occurring in the supersymmetry algebra \eqref{alg}, cf. equation \eqref{symm}. As a consequence, this product also influences the symmetry of $M$. The reason is that the anticommutators $\acom{Q^i_\ua}{Q^j_\ub}=\acom{Q^j_\ub}{Q^i_\ua}$ are symmetric under the interchange $i,\ua\leftrightarrow j,\ub$. Hence, when $\gam^a\IC$ are symmetric matrices, $M$ is symmetric, and, conversely, $M$ is antisymmetric when $\gam^a\IC$ are antisymmetric matrices (with no loss of generality). 
The matrix $M$ thus has the following symmetry property:
 \begin{align}
  M^{ji}=\epsilon\,\eta\,M^{ij}.
	\label{Msymm} 
 \end{align}
As a consequence, in the cases $\epsilon\,\eta=-1$ one needs at least two sets of supersymmetries ($\Ns\geq 2$) in order that the supersymmetry algebra \eqref{alg} has a nontrivial anticommutator $\acom{Q^i_\ua}{Q^j_\ub}$. Table \eqref{table2} shows that a choice with $\epsilon\,\eta=1$ is available in $0,\dots,4$ modulo 8 dimensions whereas it is not available in $5,6,7$ modulo 8 dimensions. 
 
We add a remark on Majorana-Weyl and symplectic Majorana-Weyl supersymmetries in even dimensions $\Dim=2k$.
In these cases the Majorana condition \eqref{MSusy1} or the symplectic Majorana condition \eqref{MSusy9} are imposed on Weyl supersymmetry generators denoted by $Q_+^{i_+}$, $Q_-^{i_-}$ where the subscripts $\pm$ indicate the chirality of the respective generator. The numbers of supersymmetry generators with positive and negative chirality are denoted by $\Ns_+$ and $\Ns_-$ respectively,
\begin{align}
  \Dim=2k:\ \quad &\GAM_\ua{}^\ub Q_{+\ub}^{i_+}=Q_{+\ua}^{i_+}\, ,\ i_+=1,\dots,\Ns_+\, ,\notag\\
  &\GAM_\ua{}^\ub Q_{-\ub}^{i_-}=-Q_{-\ua}^{i_-}\, ,\ i_-=1,\dots,\Ns_-\, .
	\label{MSusy12} 
\end{align}
A supersymmetry algebra with $\Ns_+$ Majorana-Weyl or symplectic Majorana-Weyl supersymmetry generators $Q_+^{i_+}$ and $\Ns_-$ Majorana-Weyl or symplectic Majorana-Weyl supersymmetry generators $Q_-^{i_-}$ is denoted $(\Ns_+,\Ns_-)$-supersymmetry algebra.
The above equations \eqref{MSusy1} to \eqref{MSusy11} apply to these supersymmetry algebras with 
\begin{align}
  \Ns=\Ns_++\Ns_-\, ,
	\label{MSusy13} 
\end{align}
i.e., the index $i$ runs in these cases over all sets of Majorana-Weyl or symplectic Majorana-Weyl supersymmetry generators (with positive and negative chirality).

In this context we remark that in dimensions $\Dim=4k+2$  the anticommutators of supersymmetry generators of different chiralities vanish. The latter holds because one has $Q_\pm^{i_\pm}=\half (\unit\pm\GAM)Q_\pm^{i_\pm}$ which implies
\begin{align}
  \acom{Q^{i_+}_{+\ua}}{Q^{j_-}_{-\ub}}&=\quart (\unit+\GAM)_\ua{}^\ug(\unit-\GAM)_\ub{}^\ud
  \acom{Q^{i_+}_{+\ug}}{Q^{j_-}_{-\ud}}\notag\\
  &=\quart (\unit+\GAM)_\ua{}^\ug(\unit-\GAM)_\ub{}^\ud M^{i_+j_-}(\gam^a\IC)_{\ug\ud} P_a\notag\\
  &=\quart M^{i_+j_-}((\unit+\GAM)\gam^a\IC(\unit-\GAM^\top))_{\ua\ub}P_a\notag\\
  &=\quart M^{i_+j_-}((\unit+\GAM)\gam^a\IC(\unit- (-1)^{\Dim/2}\CC\GAM \IC))_{\ua\ub}P_a\notag\\
  &=\quart M^{i_+j_-}((\unit+\GAM)\gam^a(\unit- (-1)^{\Dim/2}\GAM)\IC)_{\ua\ub}P_a\notag\\
  &=\quart M^{i_+j_-}((\unit+\GAM)(\unit+ (-1)^{\Dim/2}\GAM)\gam^a\IC)_{\ua\ub}P_a\notag\\
  &=\quart M^{i_+j_-}(1+ (-1)^{\Dim/2})((\unit+\GAM)\gam^a\IC)_{\ua\ub}P_a
	\label{MSusy14} 
\end{align}
where we used equations \eqref{GAM1} and \eqref{GAMa}.

Analogous results for $\acom{Q^{i_+}_{+\ua}}{Q^{j_+}_{+\ub}}$ and $\acom{Q^{i_-}_{-\ua}}{Q^{j_-}_{-\ub}}$ show that in dimensions $\Dim=4k$ the anticommutators of supersymmetry generators with the same chirality vanish.
Hence, we may impose, with no loss of generality,
\begin{equation}
  \Dim\ \mbox{mod}\ 4=2: \ M^{i_+j_-}=M^{j_-i_+}=0;\quad
  \Dim\ \mbox{mod}\ 4=0:\ M^{i_+j_+}=M^{i_-j_-}=0.
	\label{MSusy15} 
\end{equation}
In particular, in order that the supersymmetry algebra \eqref{alg} has a nontrivial anticommutator $\acom{Q^i_\ua}{Q^j_\ub}$ in a case $\Dim=4k$ with Majorana-Weyl or symplectic Majorana-Weyl supersymmetries, there must be at least two such supersymmetries with different chirality, i.e. one needs $\Ns_+\geq 1$ and $\Ns_-\geq 1$.

\subsection{Coboundary operators}\label{Ms}

For Majorana type supersymmetry generators $Q^i_\ua$ fulfilling equations \eqref{MSusy1} or \eqref{MSusy9} we impose on the corresponding supersymmetry ghosts $\xi^\ua_i$ according to equations \eqref{MS} or \eqref{SMS1a}, respectively:
 \begin{align}
  \mbox{Majorana supersymmetries:}\quad &\xi^{*i\,\oa}=\xi_i^\ub\, B^{-1}{}_\ub{}^\oa,
  \label{Ms1} \\
  \mbox{symplectic Majorana supersymmetries:}\quad &\xi^{*i\,\oa}=-\xi_j^\ub\, B^{-1}{}_\ub{}^\oa\,\Omega^{*ji},
	\label{Ms2} 
 \end{align}
where
 \begin{align}
  \xi^{*i\,\oa}:=(\xi^\ua_i)^*.
	\label{Ms3} 
 \end{align}
Equations \eqref{Ms1}, \eqref{MSusy1}, \eqref{MSusy5} (for Majorana supersymmetries) and \eqref{Ms2}, \eqref{MSusy9}, \eqref{MSusy11} (for symplectic Majorana supersymmetries) imply by a computation analogous to \eqref{MSusy3}:
 \begin{align}
  (\xi_i^\ua  \xi_j^\ub M^{ij}(\gam^a\IC)_{\ua\ub})^*=-\xi_i^\ua  \xi_j^\ub M^{ij}(\gam^a\IC)_{\ua\ub}\ .
	\label{Ms4} 
 \end{align}
Owing to equation \eqref{gp12}, this implies that the coboundary operator $\fb$ is real for supersymmetry ghosts subject to the Majorana condition \eqref{Ms1} or \eqref{Ms2} and the $M^{ij}$ satisfying equation \eqref{MSusy5} or \eqref{MSusy11} respectively:
 \begin{align}
  \fb=(\fb)^* .
	\label{Ms5} 
 \end{align}
Furthermore one easily deduces from equations \eqref{MSusy1}, \eqref{MSusy9}, \eqref{Ms1} and \eqref{Ms2} that the part $\xi_i^\ua Q^i_\ua$ of $\brs$ is real for Majorana type supersymmetry generators:
 \begin{align}
  &\mbox{Majorana supersymmetries:}\notag\\
  & (\xi_i^\ua Q^i_\ua)^*=\xi^{*i\,\oa}Q^*_{i\,\oa}=\xi_i^\ub\, B^{-1}{}_\ub{}^\oa B_\oa{}^\ug Q^i_\ug=\xi_i^\ub Q^i_\ub\, ,
  \label{Ms6} \\
  &\mbox{symplectic Majorana supersymmetries:}\notag\\
  & (\xi_i^\ua Q^i_\ua)^*=\xi^{*i\,\oa}Q^*_{i\,\oa}
  =-\xi_j^\ub\, B^{-1}{}_\ub{}^\oa\,\Omega^{*ji}\Omega_{ik}B_\oa{}^\ug Q^k_\ug
  =\xi_j^\ub Q^j_\ub\, . 
	\label{Ms7} 
 \end{align}
Owing to \eqref{gp11}, the part $c^a P_a$ of the coboundary operator $\brs$ is real too. Hence, for supersymmetry generators and ghosts of Majorana type defined as above, the coboundary operator $\brs$ is real,
 \begin{align}
  \brs=(\brs)^* .
	\label{Ms8}
 \end{align}
Majorana-Weyl and symplectic Majorana-Weyl supersymmetry ghosts in even dimensions corresponding to a supersymmetry generator $Q^{i_+}_{+}$ are denoted by $\xi^{+}_{i_+}$.
Accordingly Majorana-Weyl and symplectic Majorana-Weyl supersymmetry ghosts in even dimensions corresponding to a supersymmetry generator $Q^{i_-}_{-}$ are denoted by $\xi^{-}_{i_-}$. These spinors fulfill
\begin{align}
  \xi^{+\ub}_{i_+}\GAM_\ub{}^\ua=\xi^{+\ua}_{i_+}\, ,\quad \xi^{-\ub}_{i_-}\GAM_\ub{}^\ua=-\xi^{-\ua}_{i_-}\, .
\label{Ms9} 
\end{align} 

\section{\texorpdfstring{General structure of $H(\fb)$ and $H(\brs)$}{General structure of H(gh) and H(s)}}\label{general}

\subsection{\texorpdfstring{General structure of $H(\fb)$}{General structure of H(gh)}}\label{general1}

There are two obvious degrees which can be used to structure $H(\fb)$. The first is the degree of homogeneity in the translation ghosts, called "\cdeg" in the following. The second is the degree of homogeneity in the supersymmetry ghosts, called "\xdeg" in the following. 
We denote by $\Om^{p,n}$ the subspace of $\Om$ containing the members of $\Om$ with \cdeg~$p$ and \xdeg~$n$,
 \begin{align}
  \Om^{p,n}=\{\om^{p,n}\in\Om\,|\,N_c\,\om^{p,n}=p\,\om^{p,n},\ N_\xi\,\om^{p,n}=n\,\om^{p,n}\}
	\label{Ompn} 
 \end{align}
where $N_c$ and $N_\xi$ are the counting operators for the translation ghosts and the supersymmetry ghosts respectively,
\begin{align}
N_c=c^a\,\dds{c^a}\, ,\quad N_\xi=\xi^\ua_i\,\dds{\xi^\ua_i}\, .
\label{gen4} 
\end{align}
Hence, explicitly a member $\om^{p,n}$ of $\Om^{p,n}$ takes the form
\[
\om^{p,n}=c^{a_1}\dots c^{a_p}\,\xi^{\ua_1}_{i_1}\dots\ \xi^{\ua_n}_{i_n}\, 
  f^{i_1\dots i_n}_{\ua_1\dots\ua_n a_1\dots a_p}(\phi)
\]
where the coefficients $f^{i_1\dots i_n}_{\ua_1\dots\ua_n a_1\dots a_p}(\phi)$ depend on elements of the particular representation space $\rep$. 

As the translation ghosts $c^a$ anticommute, a member of $\Om^{p,n}$ is reminiscent of an ordinary differential $p$-form, with the translation ghosts playing the part of differentials $dx$. In particular, the \cdeg\ ranges from $p=0$ to $p=\Dim$ in $\Dim$ dimensions. In contrast, the \xdeg\ is not bounded from above.

The space $\Om$ in which the supersymmetry algebra cohomology is computed is thus the direct sum of the subspaces $\Om^{p,n}$ (with $0\leq p\leq\Dim$ and $n\geq 0$),
 \begin{align}
  \Om=\bigoplus_{p,n}\Om^{p,n}.
	\label{Om} 
 \end{align}

$\fb$ decrements the \cdeg\ by one unit and increments the \xdeg\ by two units, i.e., it
maps the members of $\Om^{p,n}$ for $p>0$ to members of $\Om^{p-1,n+2}$ and the members of $\Om^{0,n}$ to zero,
 \begin{align}
  \fb:\left\{\begin{array}{ll}\Om^{p,n}\longrightarrow \Om^{p-1,n+2} & \mbox{if\ $0<p\leq\Dim$,}\\
  \Om^{0,n}\longrightarrow 0.& \end{array}\right.
	\label{map} 
 \end{align}
Explicitly, $\fb$ acts on a member of $\Om^{p,n}$ ($0<p\leq\Dim$) according to
 \begin{align} 
  &\fb (c^{a_1}\dots c^{a_p}\,
  \xi^{\ua_1}_{i_1}\dots\ \xi^{\ua_n}_{i_n}\, f^{i_1\dots i_n}_{\ua_1\dots\ua_n a_1\dots a_p}(\phi))\notag\\
  &=\half\sum_{k=1}^p (-1)^{k}c^{a_1}\dots\widehat{c^{a_k}}\dots c^{a_p}\,
  \xi^{\ua_1}_{i_1}\dots\ \xi^{\ua_n}_{i_n}\xi^\ua_i\xi^\ub_j\, M^{ij}(\gam^{a_k} \IC)_{\ua\ub}\,
  f^{i_1\dots i_n}_{\ua_1\dots\ua_n a_1\dots a_p}(\phi)\label{map2} 
 \end{align}
where $\widehat{c^{a_k}}$ denotes omission of $c^{a_k}$.

$H(\fb)$ is thus the direct sum of cohomology groups $H^{p,n}(\fb)$ which denote the cohomology of $\fb$ in $\Om^{p,n}$ respectively:
 \begin{align}
  H(\fb)=\bigoplus_{p,n}H^{p,n}(\fb),\quad H^{p,n}(\fb)=H(\fb,\Om^{p,n}).
	\label{gen3} 
 \end{align}
 
$H(\fb)$ can be obtained from $\Hg(\fb)$ as follows. $\Hg(\fb)$ is represented by a set $\{P_1(c,\xi),P_2(c,\xi),\dots\}$ of primitive elements $P_A(c,\xi)\in\Omg$ numbered by $A=1,2,\dots$, i.e. every cocycle in $\Hg(\fb)$ is in $\Hg(\fb)$ equivalent to a complex linear combination of the $P_A(c,\xi)$ and no nonvanishing such linear combination is a coboundary in $\Hg(\fb)$. Since $\fb$ "does not see" the elements $\phi$ of the representation space and treats them like constants, one immediately infers that any representative $\om$ of $H(\fb)$ can be written as $\om=P_A(c,\xi)f^A(\phi)$ with "coefficients" $f^A(\phi)$ depending on elements of the representation space $\rep$. However, two things should be kept in mind:
(i) in order to be a representative of $H(\fb)$, $P_A(c,\xi)f^A(\phi)$ must be in $\Om$; (ii) a complete set of inequivalent representatives of $H(\fb)$ involves usually many (in fact, typically infinitely many) different  $f^A(\phi)$, see section \ref{1D.2} for an example.

We add a few rather elementary results on $H(\fb)$. The first result is:

\begin{lemma}[$H^{p,n}(\fb)$ for $p=\Dim$]\label{lem1gen}\quad \\
If $\fb\neq 0$, all cohomology groups $H^{\Dim,n}(\fb)$ vanish trivially,
\begin{align}
\forall n:\quad\fb\om^{\Dim,n}=0,\ \om^{\Dim,n}\in\Om^{\Dim,n}\ \Leftrightarrow\ \om^{\Dim,n}=0.
\label{gen1} 
\end{align}
\end{lemma}

{\bf Proof:} Every $\om^{\Dim,n}\in\Om^{\Dim,n}$ can be written as $\om^{\Dim,n}=V(c)p^n(\xi,\phi)$ where $V(c)=c^1\dots c^\Dim$ denotes the product of all translation ghosts and $p^n(\xi,\phi)$ depends on the supersymmetry ghosts and on the elements of the representation space $\rep$ and has \xdeg\ $n$. 
This yields $\fb \om^{\Dim,n}=-\half M^{ij}(\gam^a \IC)_{\ua\ub}\,\xi^\ua_i\xi^\ub_j\,(\6V(c)/\6c^a) p^n(\xi,\phi)$. $\6V(c)/\6c^1$, \dots , $\6V(c)/\6c^\Dim$ are linearly independent as $\6V(c)/\6c^a$ is proportional to the product of all translation ghosts other than $c^a$. Hence, $\fb \om^{\Dim,n}=0$ imposes $M^{ij}(\gam^a \IC)_{\ua\ub}\,\xi^\ua_i\xi^\ub_j\,p^n(\xi,\phi)=0$ for all values of $a$. This implies $p^n(\xi,\phi)=0$ since at least one of the polynomials $M^{ij}(\gam^a \IC)_{\ua\ub}\,\xi^\ua_i\xi^\ub_j$ is nonzero if $\fb\neq 0$. Hence, $\fb \om^{\Dim,n}=0$ implies $p^n(\xi,\phi)=0$ and thus $\om^{\Dim,n}=0$ if $\fb\neq 0$. This proves lemma \ref{lem1gen} as the implication $\Leftarrow$ in equation \eqref{gen1} is trivial. \QED
\\
\\
{\bf Remark:}
Lemma \ref{lem1gen} applies to all dimensions, all signatures, all numbers of supersymmetries and all representations 
of the respective supersymmetry algebra \eqref{alg}. In particular it thus applies also for trivial representations of any supersymmetry algebra \eqref{alg} and thus to $\Hg(\fb)$. Hence, the cohomology groups $\Hg^{\Dim,n}(\fb)$ vanish for all $n$.

Another universal and obvious result concerning $H(\fb)$ is that the cohomology group $H^{0,1}(\fb)$ is represented by complex linear combinations of the supersymmetry ghosts $\xi^\ua_i$ (with coefficients that may depend on elements of the representation space $\rep$), for any dimension $\Dim$, any signature $(t,\Dim-t)$ and any number $\Ns$ of sets of supersymmetries. Indeed, every ghost polynomial with \cdeg\ $p=0$ is trivially $\fb$-closed owing to $\fb\xi^\ua_i=0$. Furthermore, no nonvanishing ghost polynomial in $\Om^{0,1}$ is a coboundary in $H(\fb)$ because the image of $\Om$ under $\fb$ only contains ghost polynomials with \xdeg s\ $n\geq 2$, cf. equation \eqref{map}, whereas the members of $\Om^{0,1}$ have \xdeg\ $n=1$. This yields the following elementary result valid for any particular choice of $(\Dim,t,\Ns)$:

\begin{lemma}\label{lem2gen}\quad \\
If $\fb\neq 0$, there is a number $p_0\in\{0,\ldots,\Dim-1\}$ such that all cohomology groups $H^{p,*}(\fb)=\bigoplus_n H^{p,n}(\fb)$ vanish for $p>p_0$ and $H^{p_0,n}(\fb)$ does not vanish for at least one value of $n$,
\begin{align}
\exists p_0\in\{0,\ldots,\Dim-1\}:\quad (\forall p>p_0:\ H^{p,*}(\fb)=0\ \wedge\ \exists n:\ H^{p_0,n}(\fb)\neq 0).
\label{gen1a} 
\end{align}
\end{lemma}

{\bf Remark:}
$p_0$ is the border of the \cdeg\ above which $H(\fb)$ is trivial.
It will turn out that typically $p_0$ is quite small as compared to the dimension $\Dim$ and decreases as the number $\Ns$ of sets of supersymmetries increases.

\subsection{\texorpdfstring{General structure of $H(\brs)$ and its relation to $H(\fb)$}{General structure of H(s) and its relation to H(gh)}}\label{general2}

We shall now outline how $H(\fb)$ can be used to systematically compute $H(\brs)$. 

An obvious degree to structure $H(\brs)$ is the degree of homogeneity in all the ghost variables $c^a,\xi^\ua_i$. We term this degree the \gdeg. It is the sum of the \cdeg\ and the \xdeg. The counting operator corresponding to the \gdeg\ is denoted by $\Ng$. It is the sum of the counting operators in equations \eqref{gen4},
\begin{align}
\Ng=N_c+N_\xi\,.
\label{gen5} 
\end{align}
$\brs$ increments the \gdeg\ by one unit, 
\begin{align}
\com{\Ng}{\brs}=\brs\,.
\label{gen5a} 
\end{align}
$H(\brs)$ thus decomposes into cohomology groups $H^g(\brs)$, $g\geq 0$ where $H^g(\brs)$ denotes the cohomology of $\brs$ in the subspace $\Om^g$ of $\Om$ with \gdeg\ $g$. $\Om^g$ is the direct sum of the subspaces $\Om^{p,g-p}$ with $p=0,\dots,\min\{\Dim,g\}$ where $\min\{\Dim,g\}$ denotes the minimum of $\Dim$ and $g$,
\begin{align}
&\Om=\bigoplus_g\Om^g,\quad \Om^g=\{\om^g\in\Om\,|\, \Ng\,\om^g=g\,\om^g\}=\bigoplus_{p=0}^{\min\{\Dim,g\}}\Om^{p,g-p},
\label{gen6} \\
&H(\brs)=\bigoplus_g H^g(\brs),\quad H^g(\brs)=H(\brs,\Om^{g}).
\label{gen7}
\end{align}

To relate $H(\brs)$ to $H(\fb)$, we use a decomposition according to the \cdeg. $\brs$ decomposes with respect to the \cdeg\ into three parts with \cdeg\ 1, 0 and $-1$ given by $c^a P_a$, $\xi^\ua_i Q_\ua^i$ and $\fb$ respectively,
\begin{align}
&\brs=d_c+d_\xi+\fb\,,\quad 
d_c=c^a P_a\,,\quad d_\xi=\xi^\ua_i Q_\ua^i\,,\notag \\
&\com{N_c}{d_c}=d_c\,,\quad \com{N_c}{d_\xi}=0,\quad \com{N_c}{\fb}=-\fb\,.
\label{gen8} 
\end{align}
Owing to $\brs^2=0$, the parts $d_c$, $d_\xi$ and $\fb$ are antiderivations fulfilling the algebra
\begin{align}
&(d_c)^2=0,\quad \acom{d_c}{d_\xi}=0,\quad (d_\xi)^2+\acom{d_c}{\fb}=0,\quad
\acom{d_\xi}{\fb}=0,\quad (\fb)^2=0.
\label{gen9} 
\end{align}
The part $d_c$ of $\brs$ is reminiscent of the ordinary exterior derivative on differential forms, with the translation ghosts taking the place of ordinary differentials $dx$ and the translational generators $P_a$ taking the place of partial derivatives (notice that the $P_a$ commute and, at least in this respect, they are similar to partial derivatives; in fact, in simple cases the $P_a$ actually can be represented by partial derivatives or related operators, cf. section \ref{1D.2}). $d_\xi$ is somehow an analog of $d_c$ with the supersymmetry ghosts $\xi^\ua_i$ in place of the translation ghosts $c^a$ and the supersymmetry generators $Q_\ua^i$ in place of the translational generators $P_a$. Nevertheless there are important differences between $d_\xi$ and $d_c$: the supersymmetry ghosts $\xi^\ua_i$ commute whereas the translation ghosts $c^a$ anticommute, and the supersymmetry generators $Q_\ua^i$ are antiderivations with a nontrivial algebra whereas the $P_a$ are commuting derivations. As a consequence $d_\xi$ is not a coboundary operator in $\Om$ (it does not square to zero in $\Om$), in contrast to $d_c$.

To analyse the cocycle condition $\brs\om^g=0$ in $H^g(\brs)$ we decompose a cocycle $\om^g$ into parts of definite \cdeg\ according to
\begin{align}
\om^g=\sum_{p=m}^{M}\om^{p,g-p},\quad N_c\,\om^{p,g-p}=p\,\om^{p,g-p},\ N_\xi\,\om^{p,g-p}=(g-p)\,\om^{p,g-p}
\label{gen10} 
\end{align}
where $0\leq m\leq M\leq\min\{\Dim,g\}$, and where $\om^{m,g-m}\neq 0$ and $\om^{M,g-M}\neq 0$ denote the parts with lowest \cdeg\ $m$ and highest \cdeg\ $M$ occurring in the decomposition of $\om^g$.
Accordingly, the cocycle condition $\brs\om^g=0$ in $H^g(\brs)$ decomposes into
\begin{align}
\brs\om^g=0\ \LRA\ 
\left\{\begin{array}{l}
0=\fb\om^{m,g-m},\\
0=d_\xi\om^{m,g-m}+\fb\om^{m+1,g-m-1},\\
0=d_c\om^{p,g-p}+d_\xi\om^{p+1,g-p-1}+\fb\om^{p+2,g-p-2}\ \mbox{for}\ m\leq p\leq M-2,\\
0=d_c\om^{M-1,g-M+1}+d_\xi\om^{M,g-M},\\
0=d_c\om^{M,g-M}.
\end{array}
\right.
\label{gen11}
\end{align}
The equations \eqref{gen11} will be called (supersymmetric) ladder equations henceforth. They show that the part $\om^{m,g-m}$ of lowest \cdeg\ $m$ contained in an $\brs$-cocycle $\om^g$ is a cocycle in $H^{m,g-m}(\fb)$. This originates from the fact that $\fb$ is the unique part in the decomposition \eqref{gen8} of $\brs$ which decrements the \cdeg. Therefore, to relate $H(\fb)$ and $H(\brs)$ one may analyse the following problems: which cocycles of $H(\fb)$ can be "lifted" via ladder equations \eqref{gen11} to cocycles of $H(\brs)$ and when is such a cocycle of $H(\brs)$ nontrivial? These problems can be attacked by means of well-known spectral sequence methods which have been developed to analyse similar towers of equations, sometimes termed descent equations, in the context of Yang-Mills type theories, cf. \cite{DuboisViolette:1985jb,Brandt:1989gv,DuboisViolette:1992ye,Henneaux:1998rp,Barnich:2000zw,Barnich:2000me}. We shall therefore only briefly sketch how these methods can be applied in the present context referring for details to the quoted references.

We say that a cocycle $\om^{m,g-m}$ in $H^{m,g-m}(\fb)$ 
can be lifted $k$ times if there are $\om^{m+1,g-m-1}$, \dots , $\om^{m+k,g-m-k}$ fulfilling the second to the $(k+1)$th of equations \eqref{gen11}. Lifting $\om^{m,g-m}$ in this way one or more times can be obstructed and the possible obstructions lie in $H(\fb)$. 

Indeed, according to the second equation \eqref{gen11}, $d_\xi\om^{m,g-m}$ must be exact in $H(\fb)$ (more precisely in $H^{m,g+1-m}(\fb)$) in order that $\om^{m,g-m}$ can be lifted once. Now, $d_\xi\om^{m,g-m}$ is $\fb$-closed owing to the first equation \eqref{gen11} and $\acom{d_\xi}{\fb}=0$ (which yield $\fb d_\xi\om^{m,g-m}=-d_\xi\fb\om^{m,g-m}=0$), i.e. $d_\xi\om^{m,g-m}$ is a cocycle in $H^{m,g+1-m}(\fb)$. However, when $H^{m,g+1-m}(\fb)$ does not vanish, $d_\xi\om^{m,g-m}$ might be nontrivial in $H^{m,g+1-m}(\fb)$, i.e. the latter cohomology group might obstruct a first lifting of $\om^{m,g-m}$. 

Those $\om^{m,g-m}$ that can be lifted once are actually determined by the cohomology $H(d_\xi,H^{m,*}(\fb))$, i.e. by the cohomology of $d_\xi$ in $H^{m,*}(\fb)$ (this cohomology is well-defined because $d_\xi$ is a coboundary operator in $H(\fb)$ owing to $(d_\xi)^2=-\acom{d_c}{\fb}$, cf. equations \eqref{gen9}). Indeed $\om^{m,g-m}$ is an $\fb$-cocycle by the first equation \eqref{gen11}. The second equation \eqref{gen11} thus requires that $d_\xi\om^{m,g-m}$ vanishes in $H^{m,*}(\fb)$. Hence, those $\om^{m,g-m}$ that can be lifted once are cocycles in $H(d_\xi,H^{m,*}(\fb))$. Furthermore, if $\om^{m,g-m}$ is exact in $H(d_\xi,H^{m,*}(\fb))$, i.e. if $\om^{m,g-m}=d_\xi\eta^{m,g-m-1}+\fb\eta^{m+1,g-m-2}$ for some $\fb$-closed $\eta^{m,g-m-1}\in\Om^{m,g-m-1}$ ($\fb\eta^{m,g-m-1}=0$) and some $\eta^{m+1,g-m-2}\in\Om^{m+1,g-m-2}$, then one can remove $\om^{m,g-m}$ from $\om^g$ by subtracting an $\brs$-coboundary and considering $\om^{\prime\,g}=\om^g-\brs(\eta^{m,g-m-1}+\eta^{m+1,g-m-2})$ in place of $\om^g$. $\om^{\prime\,g}$ is equivalent to $\om^g$ in $H^g(\brs)$ and does not contain terms with \cdeg s\ $p\leq m$, i.e. its decomposition \eqref{gen10} either vanishes (in that case $\om^{\prime\,g}$ and $\om^g$ are trivial in $H^g(\brs)$) or starts at some \cdeg\ $m^\prime> m$.\footnote{Accordingly one selects the representatives of cohomology classes of $H(\brs)$ such that their respective decomposition \eqref{gen10} starts at a \cdeg\ $m$ which is as high as possible.}

The second and further liftings can be discussed analogously. In particular
it is easy to see that the candidate obstructions for lifting $\om^{m,g-m}$ more than one time also lie in $H(\fb)$. Indeed, suppose $\om^{m,g-m}$ 
can be lifted $k$ times. Then the first $k+1$ equations \eqref{gen11} are fulfilled for some $\om^{m,g-m}$, \dots , $\om^{m+k,g-m-k}$. The $(k+1)$th equation \eqref{gen11} reads $0=d_c\om^{m+k-2,g-m-k+2}+d_\xi\om^{m+k-1,g-m-k+1}+\fb\om^{m+k,g-m-k}$. Applying $d_\xi$ to the latter equation one obtains, using the algebra \eqref{gen9} and the $k$th equation \eqref{gen11},  $\fb(d_c\om^{m+k-1,g-m-k+1}+d_\xi\om^{m+k,g-m-k})=0$. Hence $d_c\om^{m+k-1,g-m-k+1}+d_\xi\om^{m+k,g-m-k}$ is a cocycle in $H^{m+k,g+1-m-k}(\fb)$. In order to lift $\om^{m,g-m}$ a $(k+1)$th time, the $(k+2)$th equation \eqref{gen11} requires this cocycle to be  a coboundary in $H^{m+k,g+1-m-k}(\fb)$. Hence, the candidate obstruction for this lifting is indeed in $H(\fb)$.

These considerations reveal that a cocycle in $H(\fb)$ (with \gdeg\ $g$) whose lifting is obstructed corresponds to another cocycle in $H(\fb)$ (with \gdeg\ $g+1$) obstructing this lifting. In other words, the cocycles in $H(\fb)$ which cannot be lifted and the cocycles which obstruct the lifting of other cocylces occur pairwise.
Accordingly one can prove that there is a basis $B=B_0\oplus B_\mathrm{obstructed}\oplus B_\mathrm{obstructing}$ of representatives of $H(\fb)$ such that $B_\mathrm{obstructed}$ contains representatives of $H(\fb)$ which cannot be lifted to cocycles of $H(\brs)$, $B_\mathrm{obstructing}$ contains representatives of $H(\fb)$ which obstruct the lifting of the elements in $B_\mathrm{obstructed}$, and $B_0$ contains the remaining representatives of $H(\fb)$, which can be lifted to representatives of $H(\brs)$ providing a basis of representatives of $H(\brs)$. 

Hence, in principle $H(\brs)$ can be obtained from $H(\fb)$ by constructing $B_0$ and lifting its elements. 
In actual computations the construction of $B_0$ often has practical limitations, primarily because the basis $B$ of representatives of $H(\fb)$ usually contains infinitely many elements, or it is not even really needed because the particular problem under study does not require a complete determination of $H(\fb)$ but only part of it. In fact, often it is difficult, or not even necessary, to fully determine $B$ and decompose it according to $B=B_0\oplus B_\mathrm{obstructed}\oplus B_\mathrm{obstructing}$. In such cases one may use the ladder equations \eqref{gen11} to at least partly determine $H(\brs)$, such as in particularly interesting subspaces of $\Om$ adapted to the problem under study, or to characterize $H(\brs)$ without specifying $B$ or $B_0$ explicitly (cf. sections \ref{1D.2} and \ref{1D.3} for examples). 

The \cdeg\ $p_0$ above which $H(\fb)$ is trivial, cf. lemma \ref{lem2gen}, plays a particular role in the analysis of the ladder equations \eqref{gen11} because obstructions to the lifting of representatives of $H(\fb)$ can occur only at \cdeg s\ $p\leq p_0$. Hence the analysis of the ladder equations is nontrivial only at \cdeg s\ $p\leq p_0$. 
For instance, in cases with $p_0=0$ the whole analysis of the ladder equations "collapses" to \cdeg\ $p=0$ because the cohomology groups $H^{p,*}(\fb)$ vanish for all $p>0$ and thus only the first lifting of members of $\Om^{0,*}$ may be obstructed. The above discussion of the first lifting then implies:

\begin{lemma}[$H(\brs)\simeq H(d_\xi,H^{0,*}(\fb))$ if $p_0=0$]\label{lem3gen}\quad \\
If the cohomology groups $H^{p,*}(\fb)$ vanish for all $p>0$, $H^g(\brs)$ is isomorphic to the cohomology of $d_\xi$ in $H^{0,g}(\fb)$,
\begin{align}
H^{p,*}(\fb)=0\quad \forall p>0\quad \Then\quad H^g(\brs)\simeq H(d_\xi,H^{0,g}(\fb)).
\label{gen12} 
\end{align}
\end{lemma}

\subsection{\texorpdfstring{Relating primitive elements for different signatures}{Relating primitive elements for different signatures}}\label{sign}

There is a close relation between
supersymmetry algebras \eqref{alg} in a specific dimension $\Dim$ differing only in the signature $(t,\Dim-t)$. Accordingly there is also a close relation between the primitive elements of the corresponding supersymmetry algebra cohomologies.
This is readily realized using that for every set $\{\gam^1,\dots,\gam^\Dim\}$ of gamma-matrices for signature $(t,\Dim-t)$ there is a corresponding set $\{\gam^1_{(t=0)},\dots,\gam^\Dim_{(t=0)}\}$ of gamma-matrices for signature $(0,\Dim)$, with $\gam^a=-\Ii\gam^a_{(t=0)}$ for $a\leq t$ and $\gam^a=\gam^a_{(t=0)}$ for $a>t$.
The action of the coboundary operator $\fb$ on the translation ghosts for signature $(t,\Dim-t)$ can thus be written as
\begin{align}
  \fb c^a=\left\{\begin{array}{rl}
          \ihalf M^{ij}(\gam^a_{(t=0)} \IC)_{\ua\ub}\,\xi^\ua_i\xi^\ub_j & \mathrm{if}\ a\leq t,\\
          -\half M^{ij}(\gam^a_{(t=0)} \IC)_{\ua\ub}\,\xi^\ua_i\xi^\ub_j & \mathrm{if}\ a> t.
          \end{array}\right.
\label{sign1} 
\end{align}
Defining translation ghost variables $\7c^a$ according to
\begin{align}
  \7c^a=\left\{\begin{array}{rl}
          \Ii c^a & \mathrm{if}\ a\leq t,\\
          c^a & \mathrm{if}\ a> t
          \end{array}\right.
\label{sign2} 
\end{align}
we can thus write the action of the coboundary operator $\fb$ on the translation ghosts for any signature $(t,\Dim-t)$
in the "universal form" 
\begin{align}
  \fb \7c^a= -\half M^{ij}(\gam^a_{(t=0)} \IC)_{\ua\ub}\,\xi^\ua_i\xi^\ub_j\,.
\label{sign3} 
\end{align}
Hence, for fixed dimension $\Dim$, fixed matrices $M$ and $C$ and sets of gamma-matrices related to a fixed set $\{\gam^1_{(t=0)},\dots,\gam^\Dim_{(t=0)}\}$ as described above, the coboundary operator $\fb$ has exactly the same form (i.e., with the same numerical coefficients of the supersymmetry ghosts) for different signatures $(t,\Dim-t)$ when written in terms of the translation ghost variables $\7c^a$ and the supersymmetry ghosts $\xi^\ua_i$. Of course, this statement refers to supersymmetry algebras with the same number of independent supersymmetries and the same value of $\Ns$. As a direct consequence, the primitive elements of such supersymmetry algebras may be expressed in a "universal form" in terms of the translation ghost variables $\7c^a$ and the supersymmetry ghosts $\xi^\ua_i$. In this form they are independent of the signature $(t,\Dim-t)$ and may be rewritten in terms of the original ghost variables using \eqref{sign2}. The reality relations between the supersymmetry ghosts $\xi^\ua_i$ differ, however, in general for different signatures as these reality relations are the respective Majorana or symplectic Majorana conditions \eqref{Ms1} or \eqref{Ms2} which depend on the signature $(t,\Dim-t)$. 

In particular, the above discussion shows immediately that for fixed $\Dim$, $N$, $M$ and $C$ and sets of gamma-matrices related to a fixed set $\{\gam^1_{(t=0)},\dots,\gam^\Dim_{(t=0)}\}$ as described above, the cohomologies $\Hg(\fb)$ for signatures $(t,\Dim-t)$ and $(\Dim-t,t)$ are isomorphic owing to the corresponding properties of the spinor representations for signatures $(t,\Dim-t)$ and $(\Dim-t,t)$ (cf. section \ref{spinors7}). Furthermore, the corresponding primitive elements for signature $(\Dim-t,t)$ can be directly obtained from their counterparts for signature $(t,\Dim-t)$ and vice versa by expressing them in universal form as explained above. Hence, with no loss of generality one may restrict the computation of the cohomology groups $\Hg(\fb)$ to signatures $(t,\Dim-t)$ with $t\leq \Dim/2$ (or $t\geq \Dim/2$) in order to determine the primitive elements for the various signatures in a particular dimension $\Dim$.

\subsection{\texorpdfstring{Dependence on the spinor representation}{Dependence on the spinor representation}}\label{repdep}

In section \ref{spinors6} we have outlined the relation between equivalent spinor representations. As every supersymmetry algebra \eqref{alg} involves explicitly the gamma-matrices $\gam^a$ and an inverse charge conjugation matrix $\IC$ of a spinor representation, it is natural to ask how the cohomologies $\Hg(\fb)$, $H(\fb)$ and $H(\brs)$ for equivalent spinor representations are related to one another and whether one can express these cohomologies in a manner that does not depend on the particular spinor representation.

Supersymmetry algebras \eqref{alg} involving different equivalent spinor representations are related by equations \eqref{eq1} to \eqref{eq3} with
 \begin{align}
  Q^{\prime\, i}_\ua=R_\ua{}^\ub Q^i_\ub\, ,\quad \xi^{\prime\,\ua}_i=\xi^{\ub}_iR^{-1}{}_\ub{}^\ua.
	\label{repdep1} 
 \end{align}
The coboundary operators $\fb$ and $\brs$ are evidently invariant under such a change of the spinor represenation in the sense that
 \begin{align*}
 &(\fb)^\prime=-\half M^{ij}(\gam^{\prime\, a} \CC^{\prime\,-1})_{\ua\ub}\,
 \xi^{\prime\,\ua}_i\xi^{\prime\,\ub}_j\,\frac{\6}{\6 c^a}=
 -\half M^{ij}(\gam^a \IC)_{\ua\ub}\,\xi^\ua_i\xi^\ub_j\,\frac{\6}{\6 c^a}
 =\fb\,,\\
 &(\brs)^\prime=c^a P_a+\xi^{\prime\,\ua}_i \,Q_\ua^{\prime\,i}+(\fb)^\prime
 =c^a P_a+\xi^\ua_i \,Q_\ua^i+\fb
 =\brs\,.
 \end{align*}
An immediate consequence is that the cohomologies $\Hg(\fb)$, $H(\fb)$ and $H(\brs)$ are isomorphic to their respective "primed" counterparts defined in terms of an equivalent spinor representation. 

However, it is somewhat involved to relate the representatives of an "unprimed" and a corresponding "primed" cohomology to one another. The reason is that not only the supersymmetry generators, supersymmetry ghosts and spinorial elements of the representation space $\rep$ are related by a matrix $R$ as in equations \eqref{repdep1} and \eqref{eq3}, but in addition the gamma-matrices and the charge conjugation matrices which occur in the primed and unprimed supersymmetry algebras \eqref{alg} are also related by this matrix $R$ according to equations \eqref{eq1} and \eqref{eq2}. For this reason, in general the representatives of the cohomologies $\Hg(\fb)$, $H(\fb)$ and $H(\brs)$ for different equivalent spinor representations can not be obtained from one another solely by relating the supersymmetry ghosts and the spinorial elements of the representation space $\rep$ according to equations \eqref{repdep1} and \eqref{eq3}. In addition one has to include the change of the gamma-matrices and the charge conjugation matrix according to \eqref{eq1} and \eqref{eq2}.

An expedient way to account for this influence of the spinor representation on the representatives of the cohomologies is to express these representatives appropriately in terms of $\mathfrak{so}(t,\Dim-t)$-covariant quantities (in the case of signature $(t,\Dim-t)$) built by means of gamma-matrices and/or the charge conjugation matrix (such as $c^a\xi^\ua_i\gam_{a\,\ua}{}^\ub$, $\xi^\ua_i\xi^\ub_j\IC_{\ua\ub}$ etc.). This allows one to describe the cohomologies $\Hg(\fb)$, $H(\fb)$ and $H(\brs)$ for all equivalent spinor representations at once, and thus in a form that does not depend on a particular spinor representation.

\subsection{\texorpdfstring{Dependence on the choices of $B$ and $C$}{Dependence on the choices of B and C}}\label{BCdep}

As remarked above (cf. in particular section \ref{spinors7}) one may use different matrices $B$ for the definition of Majorana type spinors ($MW$, $M$, $SMW$ or $SM$ spinors) for a particular signature $(t,\Dim-t)$, except when $\Dim$ is even and $\Dim/2+t$ is odd. This, however, does not affect the cohomology $\Hg(\fb)$ in the following sense. The matrix $B$ only enters the respective Majorana condition \eqref{MS} or \eqref{SMS1}, \eqref{SMS1a}, i.e., it only determines which type of spinors exist and how the components of the supersymmetry ghosts are related by complex conjugation, see equations \eqref{Ms1} and \eqref{Ms2}. According to these equations one may always use the components of the supersymmetry ghosts $\xi_i$ in order to express the representatives of $\Hg(\fb)$ and never needs the components of the conjugate-complex supersymmetry ghosts $\xi^{*i}$ because the components of $\xi^{*i}$ are algebraically related to those of $\xi_i$ by \eqref{Ms1} or \eqref{Ms2}. Furthermore, the matrix $B$ does not enter the coboundary operator $\fb$. Therefore, the representatives of $\Hg(\fb)$ do not depend on the choice of $B$ when expressed solely in terms of the components of the supersymmetry ghosts $\xi_i$.

This is different for the charge conjugation matrix $\CC$ as its inverse occurs explicitly in the supersymmetry algebra \eqref{alg} and in the coboundary operator $\fb$. In particular $\CC$ determines whether the matrices $\gam_a\CC^{-1}$ are symmetric or antisymmetric, cf. equation \eqref{symm}, which is relevant to the number $\Ns$ of sets of supersymmetries which can be present. In particular, in dimensions $\Dim\Mod 8=0$ and $\Dim\Mod 8=4$ there is one choice of $\CC$ with $\eta\epsilon=1$ for which the matrices $\gam_a\CC^{-1}$ are symmetric, and another choice with $\eta\epsilon=-1$ for which they are antisymmetric, cf. table \eqref{table2}. An $\Ns=1$ supersymmetry thus exists in these dimensions only for a choice with $\eta\epsilon=1$. This has to be taken into account when relating or comparing the results for $\Hg(\fb)$ for the different choices of $\CC$ in these dimensions. A second aspect that has to be taken account of in this context is that in every even dimension there are two different options for the choice of $\CC$, cf. equation \eqref{C1} and table \eqref{table2}. The two different matrices $\CC$ are related by multiplying one of them by $\GAM$ from the right (up to a factor $\pm\Ii$ or $\pm 1$). This relation may be used to relate or compare the results for $\Hg(\fb)$ valid for the two different choices of $\CC$. In particular one may use it in order to derive the results for one of these choices from the other one.
 
\section{\texorpdfstring{Examples in $\Dim=1$ dimension}{Examples in D=1 dimension}}\label{1D}

In this section we illustrate aspects of the supersymmetry algebra cohomology for simple examples in $\Dim=1$ dimension. The examples are the determination of $\Hg(\fb)$, i.e. of the primitive elements of the supersymmetry algebra cohomology, for arbitrary numbers $\Ns$ of supersymmetries in $\Dim=1$ dimension, and of $H(\brs)$ for two particular representations of the $\Ns=1$ supersymmetry algebra in $\Dim=1$ dimension. We consider in $\Dim=1$ dimension the supersymmetry algebras
 \begin{align}
  \com{P_1}{P_1}=0,\quad \com{P_1}{Q^i}=0,\quad \acom{Q^i}{Q^j}=-\delta^{ij}P_1
	\label{1Dalg} 
 \end{align}
where the "Majorana" supersymmetry generators $Q^i$ fulfill
  \begin{align}
  (Q^i)^*=\Ii\,Q^i,\quad i=1,\dots,\Ns,
	\label{1Dalg1} 
 \end{align}
i.e., $(1+\Ii)Q^i/\sqrt{2}$ are real operators.

\subsection{\texorpdfstring{Primitive elements in $\Dim=1$ dimension}{Primitive elements in D=1 dimension}}\label{1D.1}

We shall now determine the cohomology groups $\Hg(\fb)$ for all supersymmetry algebras \eqref{1Dalg}, i.e. for all numbers $\Ns$ of supersymmetries. In other words, we determine for all 
$\Ns$ a complete set of primitive elements of the corresponding supersymmetry algebra cohomology.

The coboundary operator $\fb$ reads in these cases
 \begin{align}
  \fb=\half \delta^{ij}\,\xi_i\xi_j\,\frac{\6}{\6 c^1}
	\label{1Dfb} 
 \end{align}
where the supersymmetry ghosts fulfill
\begin{align}
(\xi_i)^*=-\Ii\,\xi_i\,,\quad i=1,\dots,\Ns .
\label{1D1} 
\end{align}
Hence, the coboundary operator $\fb$ acts on the translation ghost $c^1$ according to
 \begin{align}
  \fb c^1=\half\sum_{i=1}^{\Ns}(\xi_i)^2.
	\label{1D2} 
 \end{align}

Since there is only one translation ghost, the only \cdeg s are $p=1$ and $p=0$. 
The cohomology groups $\Hg^{1,n}(\fb)$ vanish according to lemma \ref{lem1gen}. Hence, we only need to determine the cohomology groups $\Hg^{0,n}(\fb)$, i.e. the cohomology of $\fb$ in the spaces $\Omg^{0,n}$ of polynomials in the supersymmetry ghosts with \xdeg\ $n$.

In the case $\Ns=1$, equation \eqref{1D2} reduces to $\fb c^1=\half(\xi_1)^2$ and a polynomial $\om^{0,n}\in\Omg^{0,n}$ takes the form $\om^{0,n}=(\xi_1)^na_n$ with $a_n\in\mathbb{C}$. For $n\geq 2$, $\om^{0,n}$ is a coboundary owing to $(\xi_1)^na_n=\fb(2 c^1(\xi_1)^{n-2}a_n)$. For $n\in\{0,1\}$, $\om^{0,n}$ is at most linear in $\xi_1$ and thus it is not $\fb$-exact in $\Omg$ unless it vanishes, since any coboundary depends at least quadratically on $\xi_1$ in this space. Hence, in the case $\Ns=1$ the cohomology groups $\Hg^{0,n}(\fb)$ vanish for $n\geq 2$, $\Hg^{0,1}(\fb)$ is represented by $\xi_1 a_1$ and $\Hg^{0,0}(\fb)$ is represented by $a_0$, with $a_0,a_1\in\mathbb{C}$:

\begin{lemma}[$\Hg(\fb)$ for $\Dim=1$, $\Ns=1$]\label{lem1D1}\quad \\
In the case $\Ns=1$ $\Hg(\fb)$ is represented by polynomials in the supersymmetry ghosts which are at most linear in $\xi_1$:
\begin{align}
&\fb\om=0,\ \om\in\Omg\quad\Leftrightarrow\quad \om= a_0+\xi_1 a_1+\fb\eta,\ a_0,a_1\in\mathbb{C},\ \eta\in\Omg;
\label{1D7}\\
&a_0+\xi_1 a_1= \fb\eta,\ \eta\in\Omg\quad \Leftrightarrow\quad
a_0=a_1=0.\label{1D8}
\end{align}
\end{lemma}

In the cases $\Ns>1$, we expand a polynomial $\om^{0,n}\in\Omg^{0,n}$ in powers of $\xi_1$ according to
\begin{align}
\om^{0,n}=\sum_{k=0}^K (\xi_1)^k a_k(\xi_2,\dots,\xi_\Ns),\quad K\leq n
\label{1D3} 
\end{align}
where $a_k(\xi_2,\dots,\xi_\Ns)$ is a polynomial in the supersymmetry ghosts $\xi_2$, \dots, $\xi_\Ns$ with \xdeg\ $n-k$ and $K$ denotes the highest degree in $\xi_1$ occurring in this expansion.
If $K\geq 2$ we remove the term of degree $K$ in $\xi_1$ from $\om^{0,n}$ by subtracting a coboundary:
\begin{align*}
K\geq 2:\ \om^{0,n}-\fb(2 c^1(\xi_1)^{K-2} a_K(\xi_2,\dots,\xi_\Ns))
=\sum_{k=0}^{K-1} (\xi_1)^k \4a_k(\xi_2,\dots,\xi_\Ns)
\end{align*}
where
\begin{align*}
&\4a_{K-2}(\xi_2,\dots,\xi_\Ns)=a_{K-2}(\xi_2,\dots,\xi_\Ns)-((\xi_2)^2+\dots+(\xi_\Ns)^2)a_K(\xi_2,\dots,\xi_\Ns),\\ 
&k\notin \{ K-2,K\}:\ \4a_k(\xi_2,\dots,\xi_\Ns)=a_k(\xi_2,\dots,\xi_\Ns).
\end{align*}
Then we proceed as follows: if $K-1>1$, we remove analogously the term of degree $K-1$ in $\xi_1$ by subtracting another coboundary, and continue this procedure until all the terms with degrees $k>1$ in $\xi_1$ have been removed. Hence, any polynomial $\om^{0,n}\in\Omg^{0,n}$ is at most linear in $\xi_1$ up to a coboundary in $\Hg(\fb)$. Furthermore one readily verifies that a polynomial $\om^{0,n}$ which is at most linear in $\xi_1$ is $\fb$-exact in the space of ghost polynomials if and only if it vanishes. This is seen as follows:
\begin{align*}
&a_0(\xi_2,\dots,\xi_\Ns)+\xi_1 a_1(\xi_2,\dots,\xi_\Ns)=\fb(c^1 g(\xi_1,\dots,\xi_\Ns))\\
\Leftrightarrow\ &a_0(\xi_2,\dots,\xi_\Ns)+\xi_1 a_1(\xi_2,\dots,\xi_\Ns)=\half ((\xi_1)^2+\dots+(\xi_\Ns)^2)g(\xi_1,\dots,\xi_\Ns)\\
\Leftrightarrow\ &a_0=a_1=g=0
\end{align*}
where we used that $\om^{0,n}$ is a coboundary in $\Hg(\fb)$ if and only if it equals $\fb\om^{1,n-2}$ for some $\om^{1,n-2}=c^1g(\xi_1,\dots,\xi_\Ns)$ where $g(\xi_1,\dots,\xi_\Ns)$ is a polynomial in the supersymmetry ghosts with \xdeg\ $n-2$.

We conclude:

\begin{lemma}[$\Hg(\fb)$ for $\Dim=1$ and $\Ns>1$]\label{lem1D2}\quad \\
In the cases $\Ns>1$ $\Hg(\fb)$ is represented by polynomials in the supersymmetry ghosts which are at most linear in $\xi_1$:
\begin{align}
&\fb\om=0,\ \om\in\Omg\ \Leftrightarrow\ \om= a_0(\xi_2,\dots,\xi_\Ns)+\xi_1 a_1(\xi_2,\dots,\xi_\Ns)+\fb\eta,\ \eta\in\Omg;
\label{1D9}\\
&a_0(\xi_2,\dots,\xi_\Ns)+\xi_1 a_1(\xi_2,\dots,\xi_\Ns)= \fb\eta,\ \eta\in\Omg\quad \Leftrightarrow\quad
a_0=a_1=0\label{1D10}
\end{align}
where $a_0(\xi_2,\dots,\xi_\Ns)$ and $a_1(\xi_2,\dots,\xi_\Ns)$ are polynomials in the supersymmetry ghosts $\xi_2$, \dots , $\xi_\Ns$.
\end{lemma}

\subsection{\texorpdfstring{$H(\brs)$ for $\Dim=1$, $\Ns=1$: first example (off-shell)}{H(s) for D=1, N=1: first example (off-shell)}}\label{1D.2}

We shall now illustrate the representation of a supersymmetry algebra \eqref{alg} and the determination of $H(\brs)$ by a first simple example in $\Dim=1$ dimension with $\Ns=1$ supersymmetry. Hence, this example concerns the supersymmetry algebra \eqref{1Dalg} with a single supersymmetry generator $Q^1$,
 \begin{align}
  \com{P_1}{P_1}=0,\quad \com{P_1}{Q^1}=0,\quad (Q^1)^2=-\half P_1\ .
	\label{1D11} 
 \end{align}
This algebra is represented on bosonic (i.e. Grassmann even) real variables $\Phi^{(n)}$ and fermionic (i.e. Grassmann odd) variables $\Psi^{(n)}$ fulfilling $(\Psi^{(n)})^*=\Ii\Psi^{(n)}$ where $n=0,1,2,\dots$, with $\Phi^{(0)}$ and $\Psi^{(0)}$ representing functions $\Phi(x)$ and $\Psi(x)$ of a coordinate $x$ of some one-dimensional base space, and $\Phi^{(n)}$ and $\Psi^{(n)}$ representing the $n$th order derivatives $d^n\Phi(x)/dx^n$ and $d^n\Psi(x)/dx^n$ of these functions with respect to $x$. The $\Phi^{(n)}$ and $\Psi^{(n)}$ are treated as jet coordinates of an infinite jet space $J^\infty$ associated with $\Phi(x)$ and $\Psi(x)$ and make up the representation space $\rep$ in the present example. In the jet space $J^\infty$, the derivative with respect to $x$ is represented by a total derivative operator $\6$ given by
 \begin{align}
  \6=\sum_{n\geq 0}\Big(\Phi^{(n+1)}\,\dds{\Phi^{(n)}}+\Psi^{(n+1)}\,\dds{\Psi^{(n)}}\Big).
	\label{1D12a} 
 \end{align}
$\6$ represents in this example
the translational generator $P_1$ occurring in the supersymmetry algebra \eqref{1D11},
 \begin{align}
  P_1=\6\quad\Then\quad P_1\Phi^{(n)}=\Phi^{(n+1)},\quad P_1\Psi^{(n)}=\Psi^{(n+1)},\quad n=0,1,2,\dots
	\label{1D12} 
 \end{align}

The supersymmetry generator $Q^1$ is represented according to
 \begin{align}
  Q^1\Phi^{(n)}=\Psi^{(n)},\quad Q^1\Psi^{(n)}=-\half\,\Phi^{(n+1)},\quad n=0,1,2,\dots
	\label{1D13} 
 \end{align}
It can be readily verified that equations \eqref{1D12} and \eqref{1D13} provide indeed a representation of the supersymmetry algebra \eqref{1D11}.

Hence, the coboundary operator $\brs$ is in this example given by
\begin{align}
&\brs=c^1P_1+\xi_1Q^1+\half (\xi_1)^2\,\frac{\6}{\6 c^1}\,,\notag\\
&P_1=\sum_{n\geq 0}\Big(\Phi^{(n+1)}\,\dds{\Phi^{(n)}}+\Psi^{(n+1)}\,\dds{\Psi^{(n)}}\Big),\
Q^1=\sum_{n\geq 0}\Big(\Psi^{(n)}\,\dds{\Phi^{(n)}}-\half\Phi^{(n+1)}\,\dds{\Psi^{(n)}}\Big) .
\label{1D13a}
\end{align}

We shall now determine $H(\brs)$ in the space $\Om$ of polynomials in the ghost variables $c^1$, $\xi_1$ and the jet variables $\Phi^{(n)}$, $\Psi^{(n)}$, $n=0,1,2,\dots$ 
along the lines of section \ref{general}. The ladder equations \eqref{gen11} reduce in $\Dim=1$ dimension to 
 \begin{align}
  0=\fb\om^{0,g},\quad
  0=d_\xi\om^{0,g}+\fb\om^{1,g-1},\quad
  0=d_c\om^{0,g}+d_\xi\om^{1,g-1}
	\label{1D14} 
 \end{align}
where $\om^{0,g}$ or $\om^{1,g-1}$ may vanish.

Specializing the considerations of section \ref{general} to the case $\Dim=1$, one obtains that every nontrivial cocycle of $H(\brs)$ contains a nontrivial representative $\om^{0,g}$ of $H(\fb)$. From lemma \ref{lem1D1} we infer that $H(\fb)$ is represented by $f_0([\Phi,\Psi])+\xi_1 f_1([\Phi,\Psi])$ where $f_0([\Phi,\Psi])$ and $f_1([\Phi,\Psi])$ are arbitrary polynomials in the jet variables $\Phi^{(n)}$, $\Psi^{(n)}$, $n=0,1,2,\dots$ (polynomial dependence on these jet variables is collectively denoted by $[\Phi,\Psi]$). Since $f_0([\Phi,\Psi])$ has \gdeg\ $g=0$ and $\xi_1 f_1([\Phi,\Psi])$ has \gdeg\ $g=1$,  $H^g(\brs)$ vanishes for all \gdeg s $g>1$ and the only cases that remain to be studied are those with \gdeg s $g=0$ and $g=1$. Lemma \ref{lem3gen} yields $H^0(\brs)\simeq H(d_\xi,H^{0,0}(\fb))$ and $H^1(\brs)\simeq H(d_\xi,H^{0,1}(\fb))$.

$H^0(\brs)$ is thus obtained from $H(d_\xi,H^{0,0}(\fb))$. Cocycles of $H(d_\xi,H^{0,0}(\fb))$ are polynomials $f_0([\Phi,\Psi])$ fulfilling $d_\xi f_0([\Phi,\Psi])=0$ (since coboundaries in $H^{0,0}(\fb)$ vanish). $d_\xi f_0([\Phi,\Psi])=0$ is equivalent to $Q^1f_0([\Phi,\Psi])=0$. The latter condition imposes that $f_0$ does not depend on the jet variables at all and thus that $H^0(\brs)$ is represented by constants owing to the following lemma:

\begin{lemma}[Kernel of $Q^1$ -- first example]\label{lem1D3}\quad \\
The only polynomials in the jet variables $\Phi^{(n)}$, $\Psi^{(n)}$ ($n=0,1,2,\dots$) which are annihilated by $Q^1$ are polynomials of degree 0,
\begin{align}
Q^1f_0([\Phi,\Psi])=0\quad\LRA\quad f_0=a\in\mathbb{C}.
\label{1D15}
\end{align}
\end{lemma}

{\bf Proof:} $Q^1f_0([\Phi,\Psi])=0$ implies $(Q^1)^2f_0([\Phi,\Psi])=0$ and thus $P_1f_0([\Phi,\Psi])=0$ owing to \eqref{1D11}. To analyse the latter condition, we use that $f_0([\Phi,\Psi])$ is a polynomial in the jet variables $\Phi^{(n)}$ ($n=0,1,2,\dots$). Hence, there is always a value $M$ such that $f_0([\Phi,\Psi])$ does not depend on any jet variable $\Phi^{(n)}$ with $n>M$. We may thus write
\[
f_0([\Phi,\Psi])=\sum_{k=0}^K(\Phi^{(M)})^kh_k(\Phi^{(0)},\dots,\Phi^{(M-1)},[\Psi])
\]
which implies
\[
P_1f_0([\Phi,\Psi])=\sum_{k=1}^Kk\,\Phi^{(M+1)}(\Phi^{(M)})^{k-1}h_k(\Phi^{(0)},\dots,\Phi^{(M-1)},[\Psi])+\dots
\]
where the non-written terms do not depend on the jet variable $\Phi^{(M+1)}$. Hence, $P_1f_0([\Phi,\Psi])=0$ implies that the coefficient functions $h_k(\Phi^{(0)},\dots,\Phi^{(M-1)},[\Psi])$ vanish for all $k\geq 1$, i.e., $f_0$ does not depend on $\Phi^{(M)}$ at all. This holds for any value of $M$ and thus $f_0$ actually does not depend on any of the jet variables $\Phi^{(n)}$ ($n=0,1,2,\dots$). One now repeats the arguments for the jet variables $\Psi^{(n)}$ ($n=0,1,2,\dots$) and concludes $f_0=a\in\mathbb{C}$. This proves the lemma as the implication $\Leftarrow$ in \eqref{1D15} is trivial. \QED
\\
\\
$H^1(\brs)$ is obtained from $H(d_\xi,H^{0,1}(\fb))$. $H^{0,1}(\fb)$ is represented by polynomials $\xi_1 f_1([\Phi,\Psi])$. $d_\xi(\xi_1 f_1([\Phi,\Psi]))$ vanishes in $H^{0,1}(\fb)$ for any $f_1([\Phi,\Psi])$ since $H^{0,2}(\fb)$ vanishes. Hence, there is no obstruction to complete a polynomial $\xi_1 f_1([\Phi,\Psi])$ to a cocycle of $H^1(\brs)$. Explicitly the cocycle takes the form $\xi_1f_1([\Phi,\Psi])-2 c^1Q^1f_1([\Phi,\Psi])$ because of
\begin{align}
d_\xi(\xi_1 f_1([\Phi,\Psi]))=(\xi_1)^2Q^1f_1([\Phi,\Psi])=\fb(2 c^1Q^1f_1([\Phi,\Psi])).
\label{1D16}
\end{align}

We thus conclude:

\begin{lemma}[Cocycles in $H(\brs)$]\label{lem1D4}\quad \\
The general solution of the cocycle condition in $H(\brs)$ is, up to coboundaries, $a+\xi_1f_1([\Phi,\Psi])-2 c^1Q^1f_1([\Phi,\Psi])$ where $a$ is an arbitrary complex number and $f_1([\Phi,\Psi])$ is an arbitrary polynomial in the jet variables $\Phi^{(n)}$, $\Psi^{(n)}$ ($n=0,1,2,\dots$),
\begin{align}
\brs\om=0,\ \om\in\Om\ \LRA\ \om=a+(\xi_1-2 c^1Q^1)f_1([\Phi,\Psi])+\brs\eta,\ a\in\mathbb{C},\ \eta\in\Om.
\label{1D17}
\end{align}
\end{lemma}

According to lemma \ref{lem3gen} a cocycle $(\xi_1-2 c^1Q^1)f_1([\Phi,\Psi])$ is a coboundary in $H^1(\brs)$ if and only if $\xi_1f_1([\Phi,\Psi])$ is a coboundary in $H(d_\xi,H(\fb))$, i.e. if there is some $f_0([\Phi,\Psi])$ such that $\xi_1f_1([\Phi,\Psi])=d_\xi f_0([\Phi,\Psi])$, or, equivalently, $f_1([\Phi,\Psi])=Q^1 f_0([\Phi,\Psi])$. This can be easily verified explicitly:
\begin{align*}
f_1([\Phi,\Psi])=Q^1f_0([\Phi,\Psi])\ \Then\ &\xi_1f_1([\Phi,\Psi])-2 c^1Q^1f_1([\Phi,\Psi])\\
&=\xi_1Q^1f_0([\Phi,\Psi])-2 c^1(Q^1)^2f_0([\Phi,\Psi])\\
&=\xi_1Q^1f_0([\Phi,\Psi])+ c^1P_1f_0([\Phi,\Psi])\\
&=(d_\xi+d_c)f_0([\Phi,\Psi])=\brs f_0([\Phi,\Psi]).
\end{align*}
Hence, a polynomial $\xi_1f_1([\Phi,\Psi])-2 c^1Q^1f_1([\Phi,\Psi])$ is a coboundary in $H(\brs)$ if and only if $f_1([\Phi,\Psi])$ can be written as $Q^1f_0([\Phi,\Psi])$ for some polynomial $f_0([\Phi,\Psi])$.

We have thus shown:

\begin{lemma}[$H(\brs)$ -- first example]\label{lem1D5}\quad \\
$H^0(\brs)$ is represented by complex numbers, $H^1(\brs)$ is represented by
polynomials $(\xi_1-2 c^1Q^1)f_1([\Phi,\Psi])$ with $f_1([\Phi,\Psi])\neq Q^1f_0([\Phi,\Psi])$, and the cohomology groups $H^n(\brs)$ vanish for $n\geq 2$.
\end{lemma}

Let us finally add a few remarks about how the result fits in with the analysis of the supersymmetric ladder equations outlined in section \ref{general2}. We recall that $H(\brs)$ can be obtained from $H(\fb)$ by constructing a basis $B=B_0\oplus B_\mathrm{obstructed}\oplus B_\mathrm{obstructing}$ of $H(\fb)$ where $B_\mathrm{obstructed}$ contains representatives that cannot be lifted and $B_\mathrm{obstructing}$ contains representatives that are the corresponding obstructions to the lifting of the elements of $B_\mathrm{obstructed}$. $H(\brs)$ is then represented by the lifted elements of $B_0$.

In the present case the only nontrivial cohomology groups $H^{p,n}(\fb)$ are $H^{0,0}(\fb)$ which is represented by polynomials $f_0([\Phi,\Psi])$ and $H^{0,1}(\fb)$ which is represented by polynomials $\xi_1f_1([\Phi,\Psi])$. By lemma \ref{lem1D3}, no non-constant polynomial $f_0([\Phi,\Psi])$ can be lifted, while all polynomials $\xi_1f_1([\Phi,\Psi])$ can be lifted (cf. equation \eqref{1D16}). Hence, in the present example non-constant representatives $f_0([\Phi,\Psi])$ of $H^{0,0}(\fb)$ make up the part $B_\mathrm{obstructed}$ in a decomposition $B=B_0\oplus B_\mathrm{obstructed}\oplus B_\mathrm{obstructing}$ of a basis $B$ of $H(\fb)$. The obstructions to the lifting of representatives of $H^{0,0}(\fb)$ are in $H^{0,1}(\fb)$. Hence, the part $B_\mathrm{obstructing}$ of $B$ contains polynomials $\xi_1f_1([\Phi,\Psi])$ which obstruct the lifting of polynomials $f_0([\Phi,\Psi])$, i.e., which fulfill $\xi_1f_1([\Phi,\Psi])=d_\xi f_0([\Phi,\Psi])$ and thus $f_1([\Phi,\Psi])=Q^1 f_0([\Phi,\Psi])$. Accordingly, $B_0$ contains a constant $a\in\mathbb{C}$ and polynomials $\xi_1f_1([\Phi,\Psi])$ with $f_1([\Phi,\Psi])\neq Q^1f_0([\Phi,\Psi])$ which are lifted to representatives $\xi_1f_1([\Phi,\Psi])-2 c^1Q^1f_1([\Phi,\Psi])$ of $H(\brs)$, in agreement with lemma \ref{lem1D5}.

\subsection{\texorpdfstring{$H(\brs)$ for $\Dim=1$, $\Ns=1$: second example (on-shell)}{H(s) for D=1, N=1: second example (on-shell)}}\label{1D.3}

By means of a second example we shall now illustrate how a field theoretical "on-shell" representation of a supersymmetry algebra \eqref{alg} is constructed from a BRST-formulation in the so-called antifield formalism \cite{Batalin:1981jr,Henneaux:1992ig,Gomis:1994he}. Actually we use here the extended antifield formalism \cite{Brandt:1997cz} which extends the standard antifield formalism by including global symmetries in addition to local symmetries. Furthermore we shall also show that, in the example under consideration, the on-shell supersymmetry algebra cohomology is isomorphic to the (extended) local BRST cohomology. Finally we shall compute this on-shell supersymmetry algebra cohomology and comment on its relation to the corresponding off-shell supersymmetry algebra cohomology derived in section \ref{1D.2}.

The present example extends the example discussed in section \ref{1D.2} and involves again the variables $\Phi=\Phi^{(0)}$ and $\Psi=\Psi^{(0)}$ representing functions $\Phi(x)$ and $\Psi(x)$ in $D=1$ dimension and their supersymmetry transformations \eqref{1D13}. This time, however, we shall go "on-shell", i.e. we shall take into account Euler-Lagrange equations of motion for $\Phi$ and $\Psi$ using the (extended) BRST-antifield formalism which provides an expedient tool to include equations of motion through the BRST-transformations of so-called antifields. 

The second example is based on the Lagrangian
\begin{align}
{\cal L}=\half(\6\Phi)^2+\Psi\6\Psi
\label{1D19}
\end{align}
where, as before, $\6$ represents the derivative $d/dx$ with respect to $x$ in the jet space $J^\infty$.
The Lagrangian \eqref{1D19} is globally supersymmetric, for its supersymmetry transformation $Q^1{\cal L}$ equals a total derivative,
\begin{align}
Q^1{\cal L}=\6(\half\Psi\6\Phi).
\label{1D20}
\end{align}
Hence, we can employ the extended antifield formalism in order to set up an extended BRST-differential $s$ which includes both the supersymmetry transformations and the Euler-Lagrange equations of motion for $\Phi(x)$ and $\Psi(x)$ corresponding to the Lagrangian \eqref{1D19}. This yields the following extended BRST-transformations of the global (i.e. constant) ghosts $\xi_1,c^1$, the "fields" $\Phi,\Psi$ and corresponding "antifields" $\Phi^\star,\Psi^\star$ according to
\begin{align}
&s\Phi=\xi_1\Psi+c^1\6\Phi,\notag\\
&s\Psi=-\half\,\xi_1\6\Phi+c^1\6\Psi,\notag\\
&s\Phi^\star=-\6^2\Phi-\half\,\xi_1\6\Psi^\star+c^1\6\Phi^\star,\notag\\
&s\Psi^\star=-2\6\Psi+\xi_1\Phi^\star+c^1\6\Psi^\star,\notag\\
&sc^1=\half(\xi_1)^2,\notag\\
&s\xi_1=0.
\label{1D21}
\end{align}
The Grassmann parity of an antifield is opposite to the Grassmann parity of the corresponding field. Hence, $\Phi^\star$ is Grassmann odd and $\Psi^\star$ is Grassmann even.
The BRST-transformations of "derivatives" $\6\Phi,\6^2\Phi,\dots$ of the fields and antifields are defined through prolongations of the transformations \eqref{1D21}, using $\com{s}{\6}=0$ and $\6 c^1=\6 \xi_1=0$. E.g., this gives $s\6\Phi=\6 s\Phi=\6(\xi_1\Psi+c^1\6\Phi)=\xi_1\6\Psi+c^1\6^2\Phi$. $s$ is defined as an antiderivation on functions of all these variables and squares to zero by construction,
\begin{align}
s^2=0.
\label{1D22}
\end{align}

Analogously to section \ref{1D.2} we denote by $\Phi^{(n)}=\6^n\Phi$, $\Psi^{(n)}=\6^n\Psi$, $\Phi^{\star(n)}=\6^n\Phi^\star$, $\Psi^{\star(n)}=\6^n\Psi^\star$ jet variables representing the $n$th order derivative of $\Phi(x)$, $\Psi(x)$, $\Phi^\star(x)$, $\Psi^\star(x)$ with respect to $x$, respectively.\footnote{The jet variables $\Phi^{(n)}$, $\Psi^{(n)}$, $\Phi^{\star(n)}$, $\Psi^{\star(n)}$ refer to an infinite jet space corresponding to the fields $\Phi(x)$, $\Psi(x)$, the antifields $\Phi^\star(x)$, $\Psi^\star(x)$ and all derivatives $d^n/dx^n$ thereof.} Furthermore we denote polynomials in the variables 
$c^1$, $\xi_1$, $\Phi^{(n)}$, $\Psi^{(n)}$, $\Phi^{\star(n)}$, $\Psi^{\star(n)}$ ($n=0,1,2,\dots$) by $\om(c,\xi,[\Phi,\Psi,\Phi^\star,\Psi^\star])$ and the space of these polynomials by $\Om_{c,\xi,[\Phi,\Psi,\Phi^\star,\Psi^\star]}$.
The (extended) BRST-cohomology $H(s)$ is defined as the cohomology of $s$ in $\Om_{c,\xi,[\Phi,\Psi,\Phi^\star,\Psi^\star]}$.

In order to determine $H(s)$ it is most helpful to change variables to $u^\ell$, $v^\ell$, $w^I$ ($\ell=0,1,2,\dots$, $I=1,\dots,5$):
\begin{align}
&u^{2k}=\Phi^{\star(k)},\ u^{2k+1}=\Psi^{\star(k)},\ k=0,1,2,\dots\notag\\
&v^{2k}=s\Phi^{\star(k)},\ v^{2k+1}=s\Psi^{\star(k)},\ k=0,1,2,\dots\notag\\
&w^1=c^1,\ w^2=\xi_1,\ w^3=\Phi,\ w^4=\4\Psi,\ w^5=\4\Phi^{(1)}  
\label{1D23}
\end{align}
where 
\begin{align}
\Phi=\Phi^{(0)},\quad\4\Psi=\Psi^{(0)}-\half c^1\Psi^\star,\quad \4\Phi^{(1)} =\6\Phi+\half\xi_1\Psi^\star-c^1\Phi^\star.
\label{1D24}
\end{align}
The new variables \eqref{1D23} replace one by one the original variables $c^1$, $\xi_1$, $\Phi^{(n)}$, $\Psi^{(n)}$, $\Phi^{\star(n)}$, $\Psi^{\star(n)}$ ($n=0,1,2,\dots$) such that every polynomial in the original variables can uniquely be written as a polynomial in the new variables and vice versa. Indeed the variables $u^\ell$ are the original antifield variables $\Phi^{\star(n)}$, $\Psi^{\star(n)}$ ($n=0,1,2,\dots$) and the variables $w^1$ and $w^2$ are the original ghost variables $c^1$, $\xi_1$. The remaining new variables $v^\ell$, $w^3$, $w^4$, $w^5$ replace one by one the original jet variables $\Phi^{(n)}$, $\Psi^{(n)}$ ($n=0,1,2,\dots$) owing to
\begin{align}
&v^{2k}=s\Phi^{\star(k)}=-\Phi^{(2+k)}+\dots\quad (k=0,1,2,\dots),\notag\\ 
&v^{2k+1}=s\Psi^{\star(k)}=-2\Psi^{(1+k)}+\dots\quad (k=0,1,2,\dots),\notag\\
&w^3=\Phi^{(0)},\notag\\
&w^4=\Psi^{(0)}+\dots,\notag\\
&w^5=\Phi^{(1)}+\dots
\label{1D25}
\end{align}
where non-written terms are quadratic in the original variables. Hence, $\Om_{c,\xi,[\Phi,\Psi,\Phi^\star,\Psi^\star]}$ is equal to the space $\Om_{u,v,w}$ of polynomials $\om(u,v,w)$ in the variables $u^\ell$, $v^\ell$, $w^I$ and can be written as the direct product of the space $\Om_{u,v}$ of polynomials $\om(u,v)$ in the $u^\ell$, $v^\ell$ and of the space $\Om_w$ of polynomials $\om(w)$ in the $w^I$,
\begin{align}
\Om_{c,\xi,[\Phi,\Psi,\Phi^\star,\Psi^\star]}=\Om_{u,v,w}=\Om_{u,v}\otimes \Om_w\, .
\label{1D25a}
\end{align}

The new variables \eqref{1D23} have been constructed such that the $u^\ell$ and $v^\ell$ form "BRST-doublets" $(u^\ell,v^\ell)$ with $v^\ell=su^\ell$ and that, for every $w^I$, $sw^I$ can be expressed as a polynomial $r^I(w)$ in the $w$'s. The latter holds because equations \eqref{1D21} imply
\begin{align}
s\Phi=\xi_1\4\Psi+c^1\4\Phi^{(1)},\quad s\4\Psi=-\half\xi_1\4\Phi^{(1)},\quad s\4\Phi^{(1)}=0.
\label{1D26}
\end{align}
This implies that $s$ does not lead out of the subspacees $\Om_{u,v}$ and $\Om_w$ respectively,
\begin{align}
s\,\Om_{u,v}\subset\Om_{u,v}\, ,\quad s\,\Om_w\subset\Om_w\, .
\label{1D26a}
\end{align}
Owing to equations \eqref{1D25a} and \eqref{1D26a} the BRST-cohomology $H(s)$ factorizes into the cohomologies $H(s,\Om_{u,v})$ and $H(s,\Om_w)$, i.e. into the cohomologies of $s$ in the subspaces $\Om_{u,v}$ and $\Om_w$ (K\"unneth formula):
\begin{align}
H(s)=H(s,\Om_{u,v})\otimes H(s,\Om_w).
\label{1D26b}
\end{align}
As the $u^\ell$ and $v^\ell$ form BRST-doublets,
we immediately infer by standard arguments that $H(s,\Om_{u,v})\simeq \mathbb{C}$ and thus that $H(s)\simeq H(s,\Om_w)$:

\begin{lemma}[$H(s)\simeq H(s,\Om_w)$]\label{lem1D6}\quad \\
The cohomology $H(s)$ of $s$ in $\Om_{c,\xi,[\Phi,\Psi,\Phi^\star,\Psi^\star]}=\Om_{u,v,w}$ is isomorphic to the cohomology $H(s,\Om_w)$ of $s$ in $\Om_w$ and represented by representatives of $H(s,\Om_w)$:
\begin{align}
&s\om(u,v,w)=0\quad \LRA\quad \om(u,v,w)=s\eta(u,v,w)+\4\om(w),\ s\4\om(w)=0;\label{1D27}\\
&\4\om(w)=s\eta(u,v,w)\quad \LRA\quad
\4\om(w)=s\4\eta(w).
\label{1D28}
\end{align}
\end{lemma}

{\bf Proof:} 
\eqref{1D27} is proved by means of a standard contracting homotopy technique using an antiderivation $r$ defined by
\begin{align}
r=\sum_{\ell\geq 0}u^\ell\,\dds{v^\ell}\, .
\label{1D29}
\end{align}
The anticommutator $\acom{r}{s}$ is the counting operator $N_{u,v}$ of all variables $u^\ell$ and $v^\ell$,
\begin{align}
\acom{r}{s}=N_{u,v}=\sum_{\ell\geq 0}\Big(u^\ell\,\dds{u^\ell}+v^\ell\,\dds{v^\ell}\Big).
\label{1D30}
\end{align}
This yields
\begin{align}
\om(u,v,w)-\om(0,0,w)&=\int_0^1\frac{d\tau}{\tau}\,N_{u,v}\,\om(\tau u,\tau v,w)\notag\\
&=\int_0^1\frac{d\tau}{\tau}\,\acom{r}{s}\,\om(\tau u,\tau v,w)\notag\\
&=\int_0^1\frac{d\tau}{\tau}\,rs\,\om(\tau u,\tau v,w)
+s\int_0^1\frac{d\tau}{\tau}\,r\,\om(\tau u,\tau v,w)
\label{1D31}
\end{align}
where $\om(\tau u,\tau v,w)$ arises from $\om(u,v,w)$ by replacing each variable $u^\ell$ and $v^\ell$ by $\tau u^\ell$ and $\tau v^\ell$.
\eqref{1D31} implies 
\begin{align}
s\om(u,v,w)=0\quad \Then\quad 
\om(u,v,w)=\om(0,0,w)
+s\int_0^1\frac{d\tau}{\tau}\,r\,\om(\tau u,\tau v,w),
\label{1D32}
\end{align}
which yields \eqref{1D27} with $\4\om(w)=\om(0,0,w)$. \eqref{1D28} holds as a consequence of equations \eqref{1D25a} and \eqref{1D26a}. \QED
\\
\\
According to lemma \ref{lem1D6} the BRST-cohomology $H(s)$ reduces in the present example to the cohomology $H(s,\Om_w)$ of $s$ in the space $\Om_w$ of polynomials in the five variables $c^1$, $\xi_1$, $\Phi$, $\4\Psi$, $\4\Phi^{(1)}$. As can be read off from equations \eqref{1D26}, the cohomology $H(s,\Om_w)$ is actually the supersymmetry algebra cohomology $H(\brs)$ for the supersymmetry algebra \eqref{1D11} represented on the three variables $\Phi$, $\4\Psi$, $\4\Phi^{(1)}$ according to
\begin{align}
&Q^1\Phi=\4\Psi,\quad Q^1\4\Psi=-\half\4\Phi^{(1)},\quad Q^1\4\Phi^{(1)}=0,
\label{1D33}\\
&P_1\Phi=\4\Phi^{(1)},\quad P_1\4\Psi=0,\quad P_1\4\Phi^{(1)}=0.
\label{1D34}
\end{align}
Hence, the coboundary operator $\brs$ is in this example given by
\begin{align}
&\brs=c^1P_1+\xi_1Q^1+\half (\xi_1)^2\,\frac{\6}{\6 c^1}\,,\notag\\
&P_1=\4\Phi^{(1)}\,\dds{\Phi}\,,\quad Q^1=\4\Psi\,\dds{\Phi}-\half\4\Phi^{(1)}\,\dds{\4\Psi}\, .
\label{1D34a}
\end{align}

In order to determine $H(\brs)$ ($=H(s,\Om_w)$), we employ the same strategy as in section \ref{1D.2} for determining $H(\brs)$ in the case of the off-shell representation of the supersymmetry algebra \eqref{1D11} considered there. Again we conclude from lemma \ref{lem1D1} by means of the ladder equations \eqref{1D14} that $H(\brs)$ can be nontrivial at most in \gdeg s $g=0$ and $g=1$.

In the case $g=0$, arguments analogous to those used in the text preceding lemma \ref{lem1D3} show that every nontrivial representative of $H^0(\brs)$ is a polynomial $f_0(\Phi,\4\Psi,\4\Phi^{(1)})$ satisfying $Q^1f_0(\Phi,\4\Psi,\4\Phi^{(1)})=0$. In place of lemma \ref{lem1D3} we obtain in the present example:

\begin{lemma}[Kernel of $Q^1$ -- second example]\label{lem1D7}\quad \\
A polynomial $f_0(\Phi,\4\Psi,\4\Phi^{(1)})$ in the jet variables $\Phi$, $\4\Psi$, $\4\Phi^{(1)}$ is annihilated by $Q^1$ if and only if it neither depends on $\Phi$ nor on $\4\Psi$,
\begin{align}
Q^1f_0(\Phi,\4\Psi,\4\Phi^{(1)})=0\quad\LRA\quad 
\frac{\6f_0(\Phi,\4\Psi,\4\Phi^{(1)})}{\6\Phi}
=\frac{\6f_0(\Phi,\4\Psi,\4\Phi^{(1)})}{\6\4\Psi}=0.
\label{1D35}
\end{align}
\end{lemma}

{\bf Proof:}  Since $\4\Psi$ is Grassmann odd, $f_0$ can depend at most linearly on $\4\Psi$. Hence, we have $f_0(\Phi,\4\Psi,\4\Phi^{(1)})=g_0(\Phi,\4\Phi^{(1)})+\4\Psi g_1(\Phi,\4\Phi^{(1)})$ for polynomials $g_0$ and $g_1$ in $\Phi$ and $\4\Phi^{(1)}$. This gives explicitly
\[
Q^1f_0(\Phi,\4\Psi,\4\Phi^{(1)})=
\4\Psi\,\frac{\6g_0(\Phi,\4\Phi^{(1)})}{\6\Phi}-\half\4\Phi^{(1)}g_1(\Phi,\4\Phi^{(1)}).
\]
Hence, 
$Q^1f_0(\Phi,\4\Psi,\4\Phi^{(1)})=0$ imposes $\6g_0(\Phi,\4\Phi^{(1)})/\6\Phi=0$ and $g_1(\Phi,\4\Phi^{(1)})=0$ and thus $f_0(\Phi,\4\Psi,\4\Phi^{(1)})=g_0(\4\Phi^{(1)})$. Conversely, any polynomial $g_0(\4\Phi^{(1)})$ fulfills $Q^1g_0(\4\Phi^{(1)})=0$ owing to \eqref{1D33}. \QED
\\
\\
We conclude that $H^0(\brs)$ is represented by polynomials in $\4\Phi^{(1)}$.

In the case $g=1$ one obtains by arguments analogous to those used in section \ref{1D.2} that the cocycles in $H^1(\brs)$ are 
polynomials $\xi_1f_1(\Phi,\4\Psi,\4\Phi^{(1)})-2 c^1Q^1f_1(\Phi,\4\Psi,\4\Phi^{(1)})$  because there are no obstructions to lifting polynomials $\xi_1f_1(\Phi,\4\Psi,\4\Phi^{(1)})$, and that such a cocycle is a coboundary if and only if $f_1(\Phi,\4\Psi,\4\Phi^{(1)})=Q^1f_0(\Phi,\4\Psi,\4\Phi^{(1)})$ for some $f_0(\Phi,\4\Psi,\4\Phi^{(1)})$.

This yields the following result:

\begin{lemma}[$H(\brs)$ -- second example]\label{lem1D8}\quad \\
$H^0(\brs)$ is represented by polynomials $g_0(\4\Phi^{(1)})$, $H^1(\brs)$ is represented by
polynomials $(\xi_1-2 c^1Q^1)f_1(\Phi,\4\Psi,\4\Phi^{(1)})$ with $f_1(\Phi,\4\Psi,\4\Phi^{(1)})\neq Q^1f_0(\Phi,\4\Psi,\4\Phi^{(1)})$, and the cohomology groups $H^n(\brs)$ vanish for $n\geq 2$.
\end{lemma}

{\bf Comments:} Notice that equations \eqref{1D33} and \eqref{1D34} provide indeed an on-shell version of the off-shell representation of the supersymmetry algebra \eqref{1D11} given in equations \eqref{1D12} and \eqref{1D13}. Namely the Euler-Lagrange equations of motion following from the Lagrangian \eqref{1D19} give $\6^2\Phi=0$ and $\6\Psi=0$. Hence, these equations of motion set to zero all jet variables $\Phi^{(n)}$ for $n\geq 2$ and all jet variables $\Psi^{(n)}$ for $n\geq 1$. The only jet variables that derive from $\Phi$ and $\Psi$ and survive on-shell are thus $\Phi^{(0)}$,  $\Psi^{(0)}$ and $\Phi^{(1)}$ which correspond to the new variables $\Phi$, $\4\Psi$ and $\4\Phi^{(1)}$ in equation \eqref{1D24}. The representations of $Q^1$ and $P_1$ in equations \eqref{1D33} and \eqref{1D34} are precisely the on-shell versions of the representations in equations \eqref{1D13} and \eqref{1D12} respectively, since setting to zero all $\Phi^{(n)}$ for $n\geq 2$ and all $\Psi^{(n)}$ for $n\geq 1$ in equations \eqref{1D13} and \eqref{1D12} one obtains the representations in equations \eqref{1D33} and \eqref{1D34}. 

Notice also that those
jet variables $\Phi^{(n)}$ and $\Psi^{(n)}$ which are set to zero by the equations of motion correspond to the variables $v^\ell$ in the present example, cf. equations \eqref{1D25}. The latter variables do not contribute to the BRST-cohomology according to lemma \ref{lem1D6} because they form BRST-doublets with the antifields. In this way the equations of motion are taken into account by the BRST-cohomology.

\section{Final remarks}\label{conclusion}

We have defined the supersymmetry algebra cohomology for supersymmetry algebras \eqref{alg} for all dimensions $\Dim$, all signatures $(t,\Dim-t)$ and all numbers $\Ns$ of sets of supersymmetries by means of a real coboundary operator $\brs$ in terms of Majorana or symplectic Majorana supersymmetries. Furthermore we have outlined how one may systematically analyse the supersymmetry algebra cohomology by means of supersymmetric ladder equations \eqref{gen11}, starting out from a set of primitive elements of the supersymmetry algebra cohomology, and we have illustrated this strategy for two simple field theoretical examples in $D=1$ dimensions. The first example concerns an off-shell representation of the $\Dim=1$, $\Ns=1$ supersymmetry algebra, the second example concerns a corresponding on-shell representation of this algebra.

Thereby the second example illustrates a particulary useful construction of an on-shell representation of a supersymmetry algebra \eqref{alg} in the field theoretical context by means of the (extended) BRST-antifield formalism. This approach allows one to overcome certain complications occurring typically in the context of supersymmetric field theories. Namely, in a typical supersymmetric field theoretical model the commutator algebra of the supersymmetry transformations and the translations actually closes only on-shell and/or modulo gauge transformations differing from the supersymmetry transformations and translations because the commutators of two supersymmetry transformations usually contain terms that vanish only on-shell and/or terms that contain gauge transformations differing from the supersymmetry transformations and translations. 

Hence, in a typical supersymmetric field theoretical model
the supersymmetry transformations usually do not directly provide a representation of a supersymmetry algebra \eqref{alg}. In particular, an off-shell representation of a supersymmetry algebra like the representation discussed in section \ref{1D.2} is not present in a typical supersymmetric field theoretical model. Furthermore, in field theories with local supersymmetry, such as standard supergravity models, the supersymmetry and translational ghosts are "local ghosts" depending on the points of the base space (instead of constants like in the examples discussed in section \ref{1D}) and the supersymmetry transformations involve partial derivatives of these ghosts with respect to base space coordinates. 

Nevertheless, even in presence of such complications (open algebras, local ghosts) usually there is a representation of a supersymmetry algebra \eqref{alg} and a corresponding supersymmetry algebra cohomology $H(\brs)$. Typically such a representation is an on-shell representation on appropriately defined gauge covariant tensor fields (field strenghts, curvatures, matter fields and covariant derivatives thereof), with the translational generators $P_a$ represented by gauge covariant derivatives of the tensor fields and the supersymmetry generators $Q^i_\ua$ by the linearized supersymmetry transformations of these tensor fields (with linearization in the tensor fields). The "non-closure-terms" disappear from the algebra because of the on-shell nature of the representation (which removes the on-shell vanishing terms) and the linearization (which removes the terms containing the additional gauge transformations). A representation of this type can be constructed systematically by BRST-cohomological means along the lines of \cite{Brandt:1996mh,Brandt:2001tg} using variables $u^\ell$, $v^\ell$, $w^I$ such that the $u$'s and $v$'s form BRST-doublets ($u^\ell,v^\ell=su^\ell)$ and the BRST-transformations of the $w$'s take the form $sw^I=r^I(w)$ as in the simple example discussed in section \ref{1D.3}. In this approach $\{w^I\}$ contains the tensor fields on which the supersymmetry algebra \eqref{alg} is represented and the ghost variables $c^a$ and $\xi_i^\ua$ (as well as further ghost variables corresponding to additional gauge transformations, if any). $\{u^\ell,v^\ell\}$ contains partial derivatives of local ghosts, gauge fields, antifields and on-shell vanishing field variables.

We end this paper with remarks on applications and relevance of the methods and results derived in this and follow-up papers. These remarks are mainly directed to experts in BRST-cohomological methods. 

The supersymmetry algebra cohomology $H(\brs)$ shows up and is particularly relevant in the context of algebraic renormalization \cite{Piguet:1995er}, in particular within the classification of counterterms and anomalies, and of consistent deformations \cite{Barnich:1993vg} of supersymmetric (quantum) field theories by BRST-cohomological methods. Details and examples of how $H(\brs)$ arises and can be used within a BRST-cohomological analysis of supersymmetric field theories can be found in \cite{Brandt:1996au,Brandt:2002pa}. $H(\brs)$ contributes in this context an essential part to the cohomology $H(s)$ of the (extended) BRST differential $s$ on local functions (if additional symmetries are present, such as Yang-Mills type gauge symmetries, $H(s)$ is not determined solely by $H(\brs)$ but also receives contributions from the additional symmetries). This is not surprising and, in fact, similar to the role of Lie algebra cohomology in standard (non-supersymmetric) Yang-Mills theory where Lie algebra cohomology provides directly $H(s)$ (cf. section 8 of \cite{Barnich:2000zw} for a review). 

However, there is a considerable difference concerning the relevance of $H(s)$ in supersymmetric field theories\footnote{Here and in the following discussion it is always assumed that the supersymmetry transformations are contained in $s$ in the case of supersymmetric field theories.} as compared to standard Yang-Mills theories which is worthwhile to be commented on in this context and responsible for the particular importance of $H(\brs)$ and $H(\fb)$. 

In fact many important field theoretical topics, such as the classification of counterterms, anomalies and consistent deformations, are actually not obtained directly from $H(s)$ but from the cohomology $H(s|d)$ of $s$ modulo the spacetime exterior derivative $d$ on local differential forms (cf. \cite{Piguet:1995er,Barnich:2000zw} and refs. given there). $H(s)$ and $H(s|d)$ are related by so-called descent equations for $s$ and $d$ (a double complex for $s$ and $d$), see section 9 of \cite{Barnich:2000zw} for a review.
 
The relation of $H(s)$ and $H(s|d)$ is quite involved in standard Yang-Mills theories (cf. section 11.2 of \cite{Barnich:2000zw}) but very direct in supersymmetric field theories (cf. section 3 of \cite{Brandt:2002pa} for a discussion in $\Dim=4$).
The reason for the direct relation of $H(s)$ and $H(s|d)$ in supersymmetric theories is that in this case $s$ contains a translational part because the translations occur in the commutator algebra of the supersymmetry transformations. The situation is analogous to standard gravitational theories where $H(s|d)$ is directly obtained from $H(s)$ as a consequence of the presence of the spacetime diffeomorphisms in $s$ (cf. section 6 of \cite{Barnich:1995ap}). Likewise, in supersymmetric theories $H(s|d)$ is directly obtained from $H(s)$
as all the information on $H(s|d)$ is already contained in $H(s)$.

The reason for this difference between standard Yang-Mills theory and supersymmetric or gravitational theories is that, concerning these matters, the counterpart of the supersymmetric or gravitational BRST-differential $s$ is {\em not} the Yang-Mills BRST-differential $s$ but the sum $s+d$ (the $d$-part of $s+d$ provides a spacetime translational part analogous to the translational part contained in the BRST-differential of supersymmetric and gravitational theories). 

This explains why $H(s)$ in supersymmetric and standard gravitational theories is in fact more comparable to $H(s|d)$ (actually to $H(s+d)$) than to $H(s)$ in Yang-Mills theories (the cocycle condition $(s+d)\om=0$ decomposes into descent equations for $s$ and $d$). Furthermore it implies that an analog of the Yang-Mills descent equations arises in supersymmetric or standard gravitational theories by decomposing the cocycle condition $s\om=0$ into terms of definite degree in the translation ghosts (\cdeg) as this degree is the counterpart of the differential form degree in the Yang-Mills descent equations. In the case of $H(\brs)$ this decomposition provides precisely the ladder equations \eqref{gen11} which can therefore be considered an analog of the Yang-Mills descent equations for the supersymmetry algebra cohomology.

When comparing the ladder equations with the Yang-Mills descent equations, two fundamental differences stand out: the ladder equations actually establish a {\em triple complex} instead of the double complex of the Yang-Mills descent equations and comprise the coboundary operator $\fb$ which {\em decrements} the \cdeg\ by one unit. In particular the presence of $\fb$ has no counterpart in standard Yang-Mills and gravitational theories. $\fb$ determines the primitive elements of the supersymmetry algebra cohomology and makes them appear in the "bottom elements" of the ladder equations (those elements with lowest \cdeg, denoted by $\om^{m,g-m}$ in \eqref{gen11}). 
This is analogous to standard Yang-Mills theory where the primitive elements of the Lie algebra cohomology appear in the bottom forms of the descent equations. However, in sharp contrast to the primitive elements of Lie algebra cohomology, the primitive elements of the supersymmetry algebra cohomology determine in addition directly also the \cdeg s of the bottom elements as well as those \cdeg s at which the lifting of bottom elements may get obstructed (see section \ref{general2}). Actually these features apply likewise to $H(\brs)$ and to $H(s)$ in standard supersymmetric field theories because the remainung parts of $s$ (those parts that are not in $\brs$) normally do not contain another piece that decrements the \cdeg.
This makes the primitive elements of the supersymmetry algebra cohomology particularly important and useful within the BRST-cohomological analysis of supersymmetric field theories.

\end{document}